\documentclass[sigconf]{acmart}

\AtBeginDocument{%
  }

\copyrightyear{2024}
\acmYear{2024}
\setcopyright{rightsretained}
\acmConference[CHI '24]{Proceedings of the CHI Conference on Human Factors in Computing Systems}{May 11--16, 2024}{Honolulu, HI, USA}
\acmBooktitle{Proceedings of the CHI Conference on Human Factors in Computing Systems (CHI '24), May 11--16, 2024, Honolulu, HI, USA}
\acmDOI{10.1145/3613904.3642794}
\acmISBN{979-8-4007-0330-0/24/05}

\usepackage{tabularx}
\usepackage{multirow}
\usepackage{placeins}
\usepackage{framed}
\usepackage{dblfloatfix}

\begin{document}

\title{\sysname{}: Supporting Reference Recombination for Graphic Design Ideation with Generative AI}


\author{DaEun Choi}
\email{daeun.choi@kaist.ac.kr}
\affiliation{%
  \institution{KAIST}
  \city{Daejeon}
  \country{Republic of Korea}
}

\author{Sumin Hong}
\email{17101992@seoultech.ac.kr}
\affiliation{%
  \institution{Seoul National University of Science and Technology}
  \city{Seoul}
  \country{Republic of Korea}
}

\author{Jeongeon Park}
\authornote{Work done while at KAIST.}
\email{jeongeonpark1@gmail.com}
\affiliation{%
  \institution{DGIST}
  \city{Daegu}
  \country{Republic of Korea}
}

\author{John Joon Young Chung}
\authornote{Work done before joining Midjourney.}
\email{jchung@midjourney.com}
\affiliation{%
  \institution{Midjourney}
  \city{San Francisco}
  \country{CA, USA}
}

\author{Juho Kim}
\email{juhokim@kaist.ac.kr}
\affiliation{%
  \institution{KAIST}
  \city{Daejeon}
  \country{Republic of Korea}
}

\renewcommand{\shortauthors}{DaEun Choi, Sumin Hong, Jeongeon Park, John Joon Young Chung, Juho Kim}
\newcommand{\sysname}[0]{CreativeConnect}
\newcommand{\persona}[0]{Sarah}
\newcommand{\etal}{\textit{et al.}}
\newcommand\hr{\par\vspace{-.5\ht\strutbox}\noindent\hrulefill\par}

\newcommand{\john}[1]{\textbf{\textcolor{blue}{JC: #1}}}

\begin{abstract}
  Graphic designers often get inspiration through the recombination of references.
Our formative study (N=6) reveals that graphic designers focus on conceptual keywords during this process, and want support for discovering the keywords, expanding them, and exploring diverse recombination options of them, while still having room for designers' creativity.
We propose \sysname{}, a system with generative AI pipelines that helps users discover useful elements from the reference image using keywords, recommends relevant keywords, generates diverse recombination options with user-selected keywords, and shows recombinations as sketches with text descriptions.
Our user study (N=16) showed that \sysname{} helped users discover keywords from the reference and generate multiple ideas based on them, ultimately helping users produce more design ideas with higher self-reported creativity, compared to the baseline system without generative pipelines.
While \sysname{} was shown effective in ideation, we discussed how \sysname{} can be extended to support other types of tasks in creativity support.

\end{abstract}



\begin{teaserfigure}
  \centering
  \includegraphics[width=\linewidth]{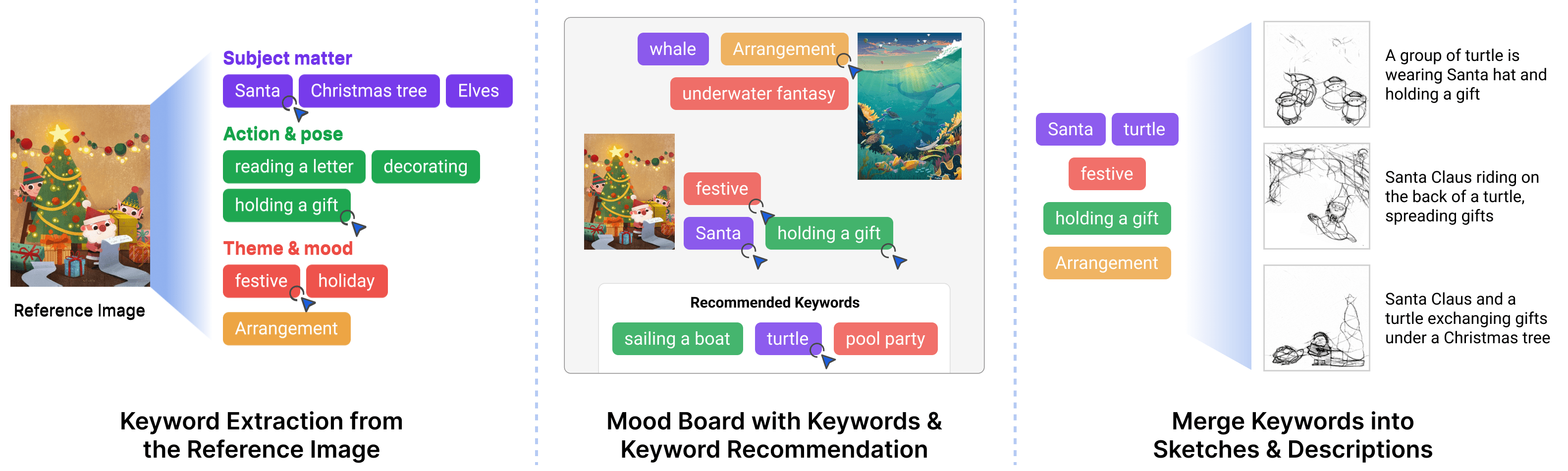}
  \caption{Three main features of \sysname{} that support reference recombination. (Left) Extraction of 4 different types of keywords from the reference image. (Center) The mood board shows the reference images with the user-selected keywords and keyword recommendations. (Right) Merging selected keywords to generate diverse recombination options and showing them as sketches and descriptions.}
  \Description{This figure briefly summarizes the three main features supported by \sysname{}. The left part depicts the keyword extraction feature, where a reference image is shown, and the system automatically extracts the keywords from the image into four categories: subject matter, action \& poses, theme \& mood, and arrangement. For example, there is Santa, a Christmas tree, and elves included in the subject matter, and decorating, holding a gift for action for \& pose, and festive, holiday for theme \& mood. The center part describes the mood board that shows those user-selected keywords along with the reference images and recommends more relevant keywords. The keywords that the user selected are shown next to the image, while some keywords (e.g., pool party) are recommended based on selected words. On the right side, the system merges keywords that the user chose. This includes Santa, turtle, festive, holding a gift, and an arrangement keyword from the specific reference image. The system merges them into descriptions (e.g., A group of turtles is wearing Santa hats and holding a gift.) and line-drawing sketches of them.}
  \label{fig:teaser}
\end{teaserfigure}

\maketitle

\section{Introduction}
References play a crucial role in creative thinking, such as graphic design, serving as valuable sources to both grasp the landscape of the existing ideas and ignite novel ones~\cite{herring2009getting, lee2010designing, ritchie2011d, muller2011leaving}. They offer diverse visual, conceptual, and functional stimuli, allowing individuals to explore various creative directions and draw lessons from established successful examples~\cite{bonnardel1999creativity}.
One effective method to generate new ideas with references is making a combination of existing examples, which is often called combinatorial creativity~\cite{campbell1960blind, simonton2003scientific, yu2011cooks, chan2016comparing}. 
In practice, this is often done through \textit{reference recombination}, which is the process of extracting the elements or aspects from multiple references, considering connections between them~\cite{herring2009getting}, and blending those to gain novel design ideas~\cite{boden1998creativity}.

However, each step of recombination requires significant effort from designers. To discover sources for recombination, designers need to dissect the references into individual elements and analyze them to determine which combinations of elements are worth mixing. Additionally, they must engage in exploratory efforts by drawing multiple sketches to find effective methods of blending those elements into a new design idea.
This takes a long time and multiple iterations, especially for those less experienced in the design process, as they have difficulty identifying various factors from references and integrating references from disparate domains compared to professionals~\cite{bonnardel2005towards}.

Previous research has provided support for these individual steps.
Several approaches have been proposed~\cite{kang2021metamap, ivanov2022moodcubes, hope2022scaling} to decompose the references or show connections between them, aiding users in identifying the sources for recombination. However, these approaches do not guide how to incorporate extracted elements into a design. Also, many approaches have attempted to help users blend different concepts or images into a novel one~\cite{chilton2019visiblends, chilton2019visifit, zhao2020iconate, wang2023popblends}. However, these approaches primarily emphasize generating precise combinations that effectively incorporate all elements harmoniously rather than aiming to produce diverse combinations for creative exploratory purposes. Another thread of research focuses on searching by genetic recombination~\cite{kim2022mixplorer, chouchoulas2007design, turrin2011design, zaman2015gem}. Still, these techniques focused on widening the range of the design exploration rather than offering inspiration on effectively combining specific design elements.

Through a formative study with six early-stage graphic designers and design students, we aimed to understand the process of reference recombination and identify their challenges.
There were two distinctive stages of ideation: 1) conceptual ideation, which aims to convey the design topic effectively, and 2) visual ideation, which is about deciding style-wise details on top of the selected concept. We decided to focus on the conceptual one as the recombination of references tends to be more prevalent in this stage.
During the conceptual ideation, designers extracted four types of elements from the reference---subject matter, action \& pose, theme \& mood, and compositional aspects (arrangement). Then, they tried to brainstorm more elements related to the extracted ones and combine them in several ways. However, due to the high effort required to recombine them manually, they were concerned that they could not try out all possible methods.
They also mentioned that support for ideation should not be in an overly completed form as it can diminish their own input.
With these observations, we propose four design goals for a reference recombination support system: (1) enable users to effortlessly specify the four types of conceptual elements from the reference image, (2) recommend relevant elements, (3) provide many recombinations as much as possible, and (4) intentionally keep the generated output partially unfinished to foster user's creativity.

Based on the design goals, we propose \sysname{}, a system that supports the design ideation process by helping users easily extract elements from the reference images and generate a wide range of recombinations of those elements. Using \sysname{}, users can easily discover and select elements from the reference image based on the four element types and get recommendations for more relevant keywords. Once the user has chosen the keywords to combine, they get various recombination options presented as pairs of sketch images and one-line descriptions. We introduced novel pipelines with generative models to automate the extraction of keywords from images, generate recombination options, and transform them into descriptions and sketches.

We conducted a within-subjects study with 16 design students to compare \sysname{} with the baseline system, which consisted of a mood board with manual keyword notes, layout diffusion model, and ChatGPT. Results showed that \sysname{} could support both stages of the reference recombination process, discovering elements from references and generating design ideas by recombining them. Participants could also produce more design ideas in a given time and perceived that \sysname{} helped them develop more creative sketches than the baseline. They emphasized that \sysname{} was especially beneficial for getting inspirational ideas vastly different from their initial concepts.
We compared the creativity support of \sysname{} with the baseline and proposed an opportunity to design a comprehensive recombination support tool that could support a broad spectrum of design needs and situations.
We also found that the low fidelity of sketch-based output led users to imagine more and get more stimulus for their creativity. Finally, we discussed the generalizability of \sysname{} in terms of user expertise, collaborative settings, and different domains of design.

This paper presents the following contributions:
\begin{enumerate}
    \item \sysname{}, a system that supports graphic designers' ideation process by helping extraction of elements from reference images and suggesting a wide range of recombinations of those elements.
    \item Computational pipelines with generative models that extract and suggest keywords from images and generate recombinations of keywords in text descriptions and sketches.
    \item Findings from a user study (N=16) about how \sysname{} can aid designers in each step of recombination, leading to the generation of more design ideas and participants to perceive their ideas as more creative.
\end{enumerate}

\section{Related Work}
This work aims to support designers in their reference recombination process for creativity. In this section, we review previous literature on (1) how references are used in graphic design ideation, (2) how recombination is employed for creative thinking, and (3) previous generative AI approaches for creativity.

\subsection{Reference in Graphic Design Ideation}
The creative process begins by collecting relevant inspirational materials from various sources~\cite{shneiderman2000creating, eckert2000sources}. Designers leverage these collected examples to gain a comprehensive understanding of the problem space. 
As the process advances into idea generation, these compiled examples play a pivotal role in fostering creativity, igniting new ideas through analogical thinking~\cite{holyoak1996mental, goel1997design}.
Recognized as one of the most challenging phases in the entire design process, previous research on creativity-supporting tools has extensively concentrated on enhancing this ideation step~\cite{frich2019mapping}. Previous research demonstrated that designers get valuable insights and inspirations in different ways~\cite{herring2009getting}, and many studies have delved into the significance of these references in design thinking, showing their potential to stimulate creativity and innovation~\cite{bonnardel1999creativity, siangliulue2015providing}.

One of the primary approaches to support idea generation with references is to help designers see diverse references. Exploring diverse ideas is crucial in terms of preventing fixation~\cite{jansson1991design}, in which a designer becomes overly fixated on a single concept, potentially hindering creativity and innovation. Therefore, Zhang et al.~\cite{zhang2021method} have utilized a Generative Adversarial Network (GAN) for exploring diverse images, while Matejka et al.~\cite{matejka2018dream} developed the Dream Lens to assist in exploring generative 3D design solution space.

Another avenue of research is to help designers manage their inspirations drawn from references, particularly through the use of mood boards~\cite{eckert2000sources}. Prior research demonstrated that building a mood board can enhance the comprehension and interpretation of ephemeral elements in design~\cite{garner2001problem}, which is beneficial for both defining and resolving design challenges~\cite{cassidy2011mood} and ultimately leading to a boost in creativity~\cite{lucero2015funky}.
Therefore, many computational systems have been proposed to help designers to build interactive mood boards, such as Funky Wall~\cite{lucero2009interactive}, SemanticCollage~\cite{koch2020semanticcollage}, and May AI~\cite{koch2019may}.

While this paper primarily focuses on the recombination of references, we have integrated two significant insights from prior research about design references. First, we emphasize the importance of offering users diverse images to support their creative processes. Second, we have incorporated the concept of a mood board as a valuable tool for organizing references within our system.

\subsection{Recombination for Creative Thinking}
In the creative thinking process, new ideas often come through the combination of the existing examples~\cite{campbell1960blind, simonton2003scientific}. It was shown that creativity often arises from forging new associations among previously unrelated frames~\cite{koestler1964act}. This process includes two crucial components: recognizing the differences between existing concepts and blending them~\cite{NAGAI2009648, boden1998creativity}. Also, the diversity of the given examples is important for building novel associations between them during this process~\cite{mumford1991process}. Observations of designers' creative processes showed that designers often maintain multiple small components and keep employing them to generate new variations through a process akin to recombination~\cite{frich2019strategies}. Many computational systems were also proposed for building recombinations and verified to be effective in tasks such as chair design~\cite{yu2011cooks} or text-based ideation~\cite{chan2016comparing}.

One practical implementation of this concept in terms of design ideation is genetic exploration. Genetic exploration involves generating novel solutions by merging elements from preexisting designs to widen the range of references. This approach has been applied in diverse domains such as garden design~\cite{kim2022mixplorer}, 3D modeling~\cite{chouchoulas2007design, pilat2008creature}, architecture~\cite{turrin2011design}, and 2D graphics~\cite{zaman2015gem}. However, these approaches primarily aim to enrich the reference in the information-gathering stage by utilizing existing references rather than supporting designers to generate their own ideas from those recombinations in the next stage.

In recombination, it is also critical to decompose the reference and get elements that are worth combining. Therefore, several tools have been developed to facilitate this process, especially by automatically decomposing the original source and showing the fine-grained aspects.
CollageMachine~\cite{kerne2000collagemachine} decomposes websites and makes them into an interactive collage. MetaMap~\cite{kang2021metamap} provides a decomposed view of the reference image into three dimensions (semantic, color, and shape) and lets users explore more references using it. Hope~\etal{}~\cite{hope2022scaling} divides the product's information into fine-grained functional parts, allowing users to combine the inspiring part. MoodCubes~\cite{ivanov2022moodcubes} offers a new mood board experience by decomposing multimedia references into constituent elements and using it to provide suggestions for new inspirational materials. They may not, however, directly discuss the exact strategies for merging these outputs as a new design idea. On the other hand, VRicolage~\cite{stemasov2023immersive} enables users to decompose objects into different parts, motions, or colors, and mix them. However, this process was more of utilizing collected assets rather than generating a new idea from recombination.

Additionally, many previous approaches supported the process of mixing the reference images or concepts. For example, VisiBlends~\cite{chilton2019visiblends} and VisiFit~\cite{chilton2019visifit} introduced a novel pipeline to blend two objects to convey integrated meaning. ICONATE~\cite{zhao2020iconate} supports users to generate a new icon by mixing different icons, and PopBlends~\cite{wang2023popblends} automatically suggests conceptual blends of reference images. FashionQ~\cite{jeon2021fashionq} supports this blending in the domain of fashion design. Artinter~\cite{chung2023artinter} supports recombining style elements from the reference to facilitate communication.
Nevertheless, these approaches primarily focus on seamlessly merging entire references rather than breaking them down to the element level. This approach may not fully align with the creative recombination process, which often begins by identifying specific elements to combine within the provided examples. 3DALL-E~\cite{liu20233dall} presents a recombination workflow for generating a new idea, which suggests diverse low-level keywords and combines them into a prompt for text-to-image models. This approach, however, differs from our definition of reference recombination as the keywords are from LLM's understanding of the world rather than the design references.

\subsection{Generative AI Approaches for Creativity}

Before looking into AI systems for creativity, it's important to know how visual designers perceive AI for supporting their design tasks. Ko~\etal~\cite{ko2023creativeworks} looked into how graphic designers use large-scale text-to-image generation models (LTGMs) to help with their creative works
and suggest the design guidelines for building creative supporting systems using them.

Recently, diffusion-based techniques~\cite{Rombach2022latent, ramesh2022hierarchical, nichol2022glide} and CLIP embedding~\cite{radford2021clip} enable people to represent their ideas to visual materials quickly and easily using text prompting. There were also many previous approaches to incorporate inputs with additional modalities, such as layout~\cite{li2023gligen, Zheng2023layout, chen2023trainingfree} or sound~\cite{lee2023aadiff, sung2023sound}. 
Techniques to add extra conditions and styles for more granular control have been proposed as well~\cite{zhang2023controlnet, mou2023t2iadapter}. There is also a thread of research on modifying the generated image to align with user intent better, such as adding style~\cite{gatys2016styletransfer}, latent-space manipulation~\cite{Karras2019stylegan, Karras2020stylegan2}, human-prompt editing~\cite{brooks2022instructpix2pix}, and editing a specific part in generated images~\cite{Ruiz2023dreambooth, gal2022textual}.

With those novel ML techniques, the creative landscape is continuously reshaped, offering innovative solutions and enriching the artistic experience.
Promptify~\cite{brade2023promptify} stands out as an iterative prompt refine tool, letting users get closer to their intended result by clearing unintended outcomes. PromptPaint~\cite{john2023prompt} allows users to go beyond language to mix prompts to express challenging concepts, supporting the iterative shaping of the image.
On the other aspect, the interplay between humans and AI is also fast evolving. The concept that Karimi~\etal~\cite{CSP2020Karimi, RCM2019Karimi} proposes is a generative AI system that helps designers by collaborating during the design phase instead of taking over the design process. Oh~\etal~\cite{duetdraw2018Oh} and Framer~\cite{lawton2023tool} proposed a user-AI collaborative interface to allow a co-drawing experience.

While there has been a lot of research on expressing user intention to ML models accurately to get a better image or collaborate with AI during the design execution phase, it is less relevant to the ideation task of expanding the variety of ideas. Specifically, it still needs to be discovered how to design interaction with generative AI models to inspire graphic designers by recombining the references.

\section{Formative Study}
\label{sec:formativestudy}
We conducted a formative study to understand how designers recombine design references for ideation and what challenges they encounter during the process.

\subsection{Participants}
The prior research~\cite{bonnardel2005towards} suggested that the less experienced tend to encounter more challenges in getting inspiration from references and combining them. Therefore, we targeted \textit{early-stage designers} as they are expected to have an understanding of the overall design process but still struggle with many challenges in ideation through recombination compared to professional designers. We defined an early-stage designer as someone who got a design education in university or has less than 3 years of professional experience as a designer.

Six participants (6 female; age M=25.3 and SD=3.32) were recruited through an online recruitment posting. Two were professional UI/UX designers with 1 year of experience each, and one was a freelance brand designer with 3 years of experience. Three were students majoring in industrial design, with two at the graduate level and one in their fourth year of undergraduate studies. All participants reported that they had experience in at least three different graphic design projects before.

\subsection{Study Process}
The study included (1) an observation on reference searching and idea sketching and (2) a semi-structured interview. For the first part of the study, participants were asked to draw an illustration for one of three different design topics they chose: "Tourism service for kids," "Pet grooming service," or "Eco-friendly restaurant." They were first given 10 minutes to search for reference images that they wanted to use. For each reference they chose, they were asked to describe what aspects of the references they found appealing. Then, participants sketched their design ideas using their preferred method for 30 minutes. Three participants used pen and paper to sketch their ideas, while the other three used a tablet and digital drawing software. They were asked to generate at least three distinct design ideas and describe how they integrated their references into each sketch. After that, we conducted a semi-structured interview to ask about their challenges in generating multiple ideas using references.

After each study session, two authors independently coded the recombination methods the participants employed in their ideation tasks and the semi-structured interview results. The coded data were then discussed collaboratively. After conducting six studies, codes were saturated, and no further study sessions were conducted.

\subsection{Findings}

Through the observation of participants' design processes, we discovered that the recombination of different references primarily occurs during the initial stages of design ideation, with a specific focus on the conceptual aspects of the reference images rather than the visual elements. We identified four distinct categories of elements employed in this process. We also found specific challenges associated with it and observed that the system supporting this process should reserve a degree of incompleteness to encourage creativity.

\subsubsection{Early-Stage Design Ideation Focuses on Conceptual Aspects}\label{sec:conceptualideation}
All six participants said they refer to the references in two distinct stages: conceptual ideation and visual development. During the conceptual ideation stage, designers focus on elements that could effectively convey the design topic, such as objects or mood. After looking at those elements, they generated multiple drafts by combining them in several ways. On the other hand, the visual development stage revolved around adding visual details like color and texture to complete the sketches derived from the conceptual ideation. During this stage, designers often had a clear direction in their mind and referred to a specific set of references that aligned well with their chosen direction, with less emphasis on exploring different recombinations of diverse references. This aligns with findings from previous research~\cite{holinaty2021supporting}, which shows that artists engage in a spectrum of reference usage in their creative process, ranging from detailed recreation (visual development) by tracing images to interpretive inspiration for high-level components (conceptual ideation).
In summary, designers recombined references primarily for conceptual ideation, which was usually the first step of the design ideation, suggesting that a system supporting the reference recombination process should focus on how to facilitate this early-stage step.

\subsubsection{Types of Elements Used for Recombination}\label{ref:channels}
During the conceptual ideation phase, participants tried to extract specific elements from references and incorporate them into their design concepts. They employed a variety of approaches for this.
The simplest approach observed in all participants was utilizing objects in a reference in their sketch. For example, for drawing an illustration for "Tourism service for kids," one participant took an image of a paper plane from a reference to convey the image of playful children and tour service at the same time.
Five participants extracted the abstract semantic meaning or overall theme conveyed by references. For example, after looking at an image of a person holding a pamphlet and deep in thought, one participant said that the keyword "imagination" could effectively capture the concept of kids. So, they developed a design concept about children imagining various travel destinations.
Another approach observed in three out of six participants was to take the action of a character from a reference. For instance, by looking at a reference illustrating an animal and a person holding hands, a participant got the concept of children holding hands together.
Lastly, five participants referred to the composition from the reference images. For example, by looking at a reference where leaf shapes were arranged together to form a shovel, one participant came up with the idea of using multiple tree trunk shapes to represent the structure of a building.

\subsubsection{Challenges During Finding Elements}
We identified some opportunities to support the process of extracting elements from the references. There were many cases where the elements designers initially found appealing in the reference search phase differed from those they eventually utilized in their design concepts. In the interview, participants said that upon closer examination of the references, they discovered new elements of interest and incorporated them. This means that designers couldn't immediately extract elements upon viewing the reference, and it often required several examinations to uncover such elements, which was time-consuming.

Another observation was that participants often came up with new keywords based on what they had already found for further brainstorming at the element level. For instance, P3 identified "toy blocks" from one reference and "train" from another reference, then came up with the new keyword "toy train" and incorporated it into their final idea. However, this process was often more challenging than finding elements directly from the reference images.
P4 highlighted an opportunity for system support this thinking process by mentioning that ``I usually talk with others about my ideas, which leads me to discover new keywords related to the original one. Just like that, I think it would be nice if the system could recommend a new keyword to expand my current design idea.''

\subsubsection{Challenges During Recombining Elements}
After finding out the elements from the references they want to utilize in their design ideas, another challenge became apparent. While there can be numerous ways to combine these elements, participants were often frustrated as they couldn't sketch out all the possibilities to determine if they were viable.
Three out of six participants expressed anxiety about not being able to consider all possible combinations. P3 stated, ``I always feel anxious that there might be a better way, but I can't think of it.'' P6 also mentioned that ``The more options I explore, the more I become confident about my final design idea. I want some faster way to explore alternatives as much as possible.''
Four out of six participants said they rely on their imagination to envision numerous recombination possibilities within their minds, as sketching out all is too time-consuming and effortful. However, two participants expressed frustration that, although combinations seemed good in their minds, they might not come together as effectively in actual sketches.

\subsubsection{System Support should be Incomplete}
Designers tended to deliberately exclude visual details during conceptual ideation.
Participants said that when recombining the references for conceptual inspiration, they did not pay much attention to visual details, and several participants noted that they even needed to exclude those details intentionally. P2 stated, ``When combining different concepts, colors and textures often become messy, so I deliberately use the same brush for all elements.'' P3 agreed with another viewpoint by expressing concern about becoming overly fixated on frequently recurring visual details while exploring conceptual recombinations.
We also asked the participants which form would be preferred if they could get recommendations for different recombination options. Four participants mentioned that they would prefer incomplete outputs, such as a sketch or even a textual description of the idea so that they could focus on the concept itself. The main reason for this was the concern that the model would compromise their creativity or lead them to perform unintentional plagiarism.

\subsection{Design Goals}
Based on the findings of the formative study, we identified four design goals to build a system to support designers' reference recombination process during early-stage ideation.

\begin{itemize}
  \item[DG 1.]  \textbf{Facilitate Element Extraction from References.} To help users efficiently find the elements that would be used for the recombination, the system should help users discover the overlooked elements. Based on our observation, elements that users want to extract from references are (1) subject matters (e.g., objects, characters, landscapes), (2) action \& pose, (3) theme \& mood, and (4) arrangement.
  \item[DG 2.] \textbf{Suggest Diverse and Relevant Elements.} To help users explore more elements on top of what they found from the references, the system should provide some recommendations of relevant elements that users might like.
  \item[DG 3.] \textbf{Generate Diverse Recombination Options.} To help users explore diverse recombination possibilities, the system should show users a varied range of recombination options and reduce their anxiety over not considering all feasible combinations. This goal highlights the system's ability to propose combinations that users might not have considered independently.
  \item[DG 4.] \textbf{Present Recombination in an Incomplete Format.} To align with designers' preference for conceptual sketches over highly detailed artwork during the initial ideation phase, the system-generated outputs should be intentionally incomplete, such as sketches. This emphasizes the importance of allowing users to inject their own creativity into the images.
\end{itemize}

\section{\sysname{}}
\begin{figure*}
  \includegraphics[width=\textwidth]{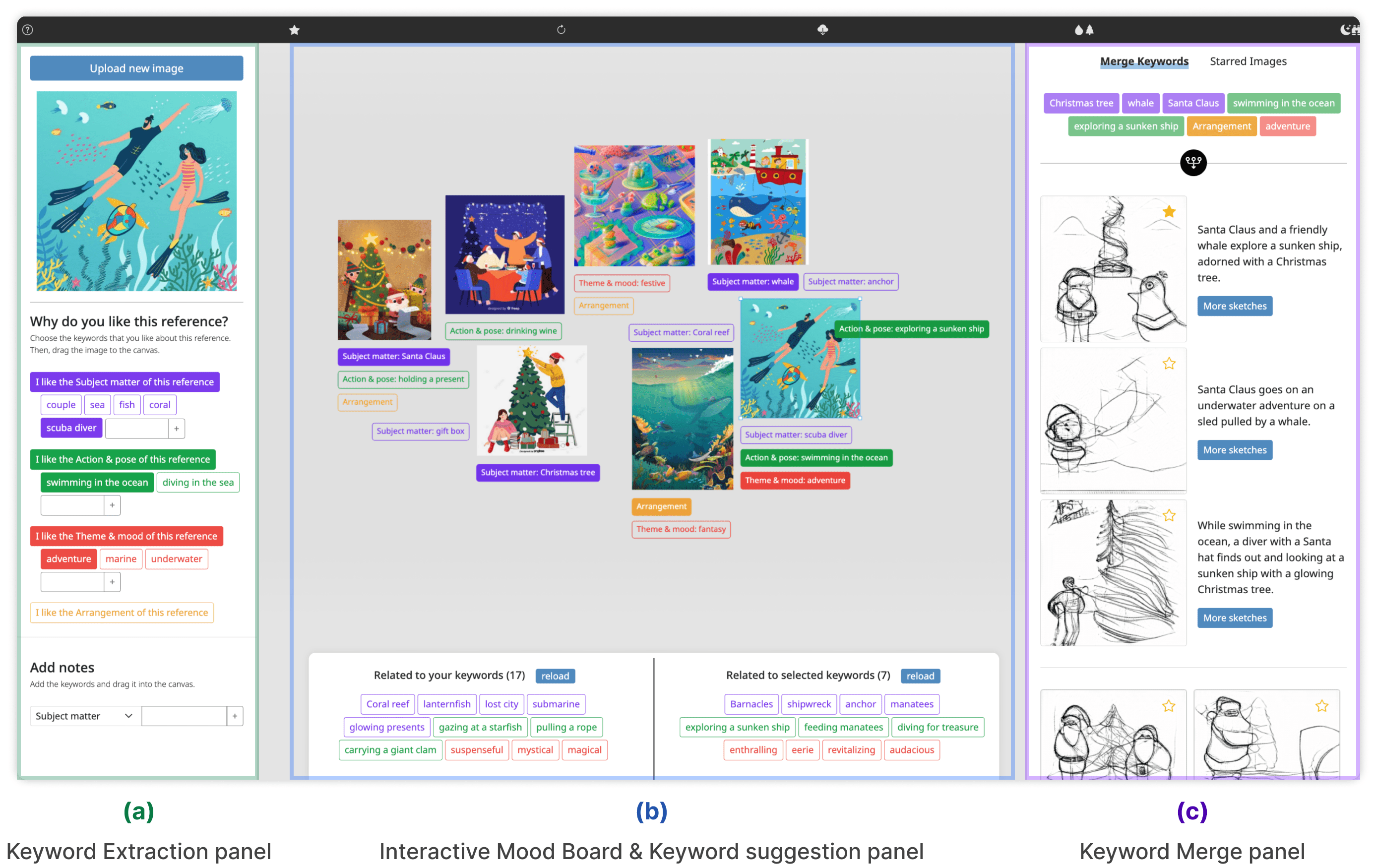}
  \caption{Screenshot of \sysname{}. (a) \textbf{Keyword Extraction Panel}: The system automatically extracts keywords in four categories (subject matter, action \& pose, theme \& mood, and arrangement) from the reference image. Users can select these keywords or add keywords manually. (b) \textbf{Interactive Mood Board with Keyword Suggestion Panel}: Users can organize the reference images along with the selected keywords. Users can import the keywords shown below, which are suggested based on all keywords on the board or the keywords that users selected on this mood board. (c) \textbf{Keyword Merge Panel}: When users select keywords they want to recombine on the mood board, the system generates sketches and their respective descriptions, including all selected keywords. Users can view more generated sketches by clicking the "More Sketches" button.}
  \Description{This is the figure of the overall \sysname{} system scene. (a) On the left, there is a keyword extraction panel. When the user uploads a reference image, the system automatically extracts keywords in four categories (subject matter, action \& pose, theme \& mood, and arrangement) and displays them here. Users can select the extracted keywords or add keywords manually. (b) In the center, there is an interactive mood board where users can organize the reference images on the mood board, and the selected keywords are placed near the corresponding image. At the bottom, there is a keyword recommendation panel where users can explore the system-recommended keyword suggestions, which are suggested based on all keywords on the board or the keywords that the user selected on this mood board. On the right side, there is a keyword merge panel. When the user selects keywords they want to recombine on the mood board, the system mixes them and generates three pairs of sketches and their respective descriptions, including all chosen keywords. There is a "More Sketches" button for each pair, which users can click to view additional sketches for the descriptions.}
  \label{fig:system}
\end{figure*}

With derived design goals, we implemented \sysname{} (Figure \ref{fig:system}), an AI-powered design tool that supports graphic designers in coming up with novel design ideas by recombining reference images in early-stage conceptual ideation.
\sysname{} mainly consists of a mood board where users can import reference images and select what they like about the reference. When the user imports a new image, the system extracts keywords according to the four categories defined in the formative study (Section ~\ref{ref:channels}) so that users can choose among them. This helps users easily discover and select keywords (DG 1). Selected keywords are then displayed on the mood board along with the images.
\sysname{} offers further keyword recommendations based on the keywords users have added to the board or their specific selections (DG 2). Also, when the user chooses a set of keywords to recombine, the system generates multiple drafts with diverse ways of combining them (DG 3). All system-generated recombination outputs are produced in line sketches with one-line descriptions so that users can further reinterpret by themselves (DG 4).

\subsection{User Scenario}
To demonstrate our system, we show how \persona{}, a junior illustration designer, uses \sysname{} to generate ideas for her design project. \persona{} recently accepted a new commission to draw an illustration for the cover of a children's book titled, "A Christmas Dinner in the Underwater World." As the given topic is an unusual combination of two themes, she struggled to get inspiration from the references and mix them to come up with ideas, so she decided to explore references with the help of \sysname{}.

\subsubsection{Getting User Inputs on the Design Reference}
\persona{} first uploads ten reference images she got from her client into \sysname{}. Looking through the references, she is intrigued by the one where two scuba divers swim with a turtle.
When she chooses the image, \sysname{} shows some keywords that can be found in the image, divided into four categories -- subject matter, action \& pose, theme \& mood, and arrangement (Figure \ref{fig:system} (a)). As she finds the scuba diver concept interesting, she clicks on the subject matter category. She finds "scuba diver" in the keyword list and clicks it. She also finds "coral reef" on the list, which she didn't recognize before. She looks at the references again and thinks coral reefs would look great in her illustration, so she clicks "coral reef" as well. Similarly, she looks through the list of the keywords in the "action \& pose" and "theme \& mood" categories and selects "swimming" and "adventure" from each list. She also likes the overall composition of the image, so she clicks its "arrangement" as well. She also works on selecting keywords that she likes on other references.

\subsubsection{Mood Board with the User-selected Keywords \& Keyword Recommendation}
As \persona{} selects the keywords she finds useful from each image, the canvas of the \sysname{} offers a dynamic mood board that shows the references with user-selected keywords, capturing her creative goal and preferences (Figure \ref{fig:system} (b)). As she freely moves the images to organize them, the selected keywords move along with the image.
By looking at the keywords, \persona{} wants to come up with additional ideas for character actions that align with the adventurous theme, similar to swimming or scuba diving. Therefore, she selects "subject matter: scuba diver", "action \& pose: swimming", and "theme \& mood: adventure" to get system recommendations with these keywords. \sysname{} shows a set of keywords, such as "action \& pose: exploring sunken ship", and "subject matter: anchor". \persona{} finds those keywords valuable, so she drags them into the mood board.

\subsubsection{Recombining Design References using Keywords}
From the set of keywords on the mood board, \persona{} now selects some keywords she wants to include in her design and uses the system to make a first draft. 
She selects "Christmas tree" and "Santa Claus" for a Christmas dinner theme, and "whale", "swimming", "exploring the sunken ship", and "adventure" for the underwater theme. She also selects the "arrangement" of one of the images with an interesting composition.

After clicking the merge button, \sysname{} generates three different drafts, each showcasing a unique and different way of incorporating these keywords (Figure \ref{fig:system} (c)). Each draft contains a one-line text description of the image concept and a sketch-style image generated based on the description and the arrangement that \persona{} selected. 
She appreciates the results as the way each draft combined keywords would be difficult to think of by herself and that all three drafts are distinct from each other. Also, the sketch format allows her to imagine further design concepts rather than fixating on the concept and details in the generated results.

Among the drafts, \persona{} finds one description interesting: "Santa Claus goes on an underwater adventure on a sled pulled by a whale." However, she feels dissatisfied with the generated sketch and presses the "More Sketches" button. Then, \sysname{} generates five more sketches with the same description but in a slightly different way. She gets some good design ideas from the new sketches and starts working on her draft.

\subsection{Technical Details}

\sysname{} was built as a web-based system with a ReactJS\footnote{https://react.dev/}-based front-end client and a Flask\footnote{https://flask.palletsprojects.com/}-based back-end server. We implemented ML pipelines for extracting the keywords from the references and merging keywords into recombinations. The technical details of these pipelines are discussed in the following sections. Some examples of outputs from the pipeline are presented in Figure ~\ref{fig:technicaloutputexample} in the Appendix.

\begin{figure*}
  \includegraphics[width=\linewidth]{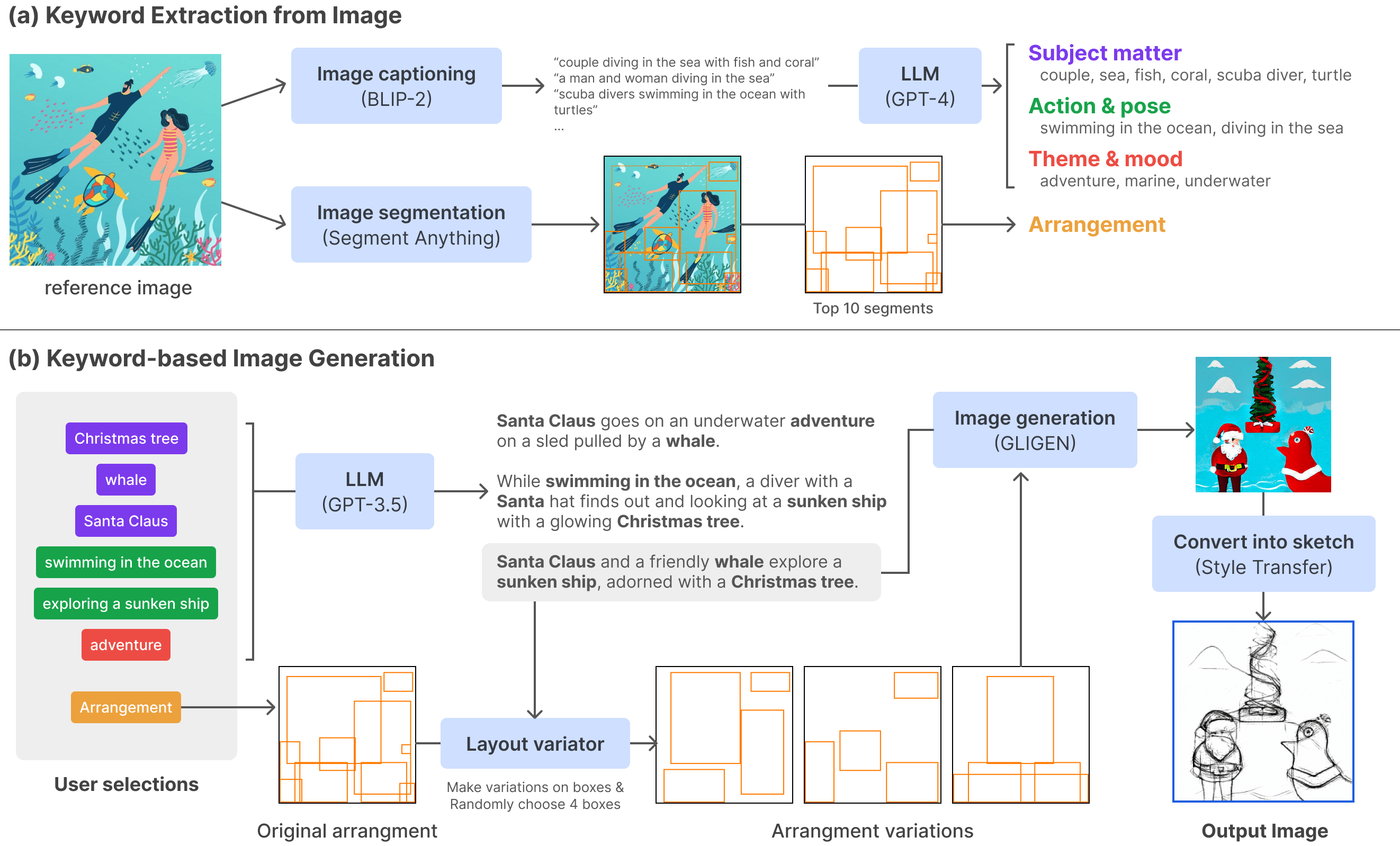}
  \caption{Technical pipeline of \sysname{}. (a) \textbf{Keyword extraction from image}: The caption generated from the image captioning model goes into the LLM to extract subject matter, action \& pose, and theme \& mood. The segmentation model is used to detect the image's arrangement. (b) \textbf{Keyword-based image generation}: the LLM generates descriptions based on the given keywords, and the layout variator generates similar arrangements. The image generation model generates the image, and the style transfer model converts this into a sketch.}
  \Description{This figure depicts the technical pipeline of \sysname{}. (a) \textbf{Keyword extraction from image}: From the input reference image, the image captioning model generates captions for the image and the LLM is used to extract subject matter, action \& pose, and theme \& mood from the captions. In parallel, the segmentation model is used to detect the arrangement. (b) \textbf{Keyword-based image generation}: the LLM first generates three descriptions using the given keywords, and the layout variator transforms the original arrangement into three similar arrangements. Then, the image generation model uses the description and layout to generate the image, and the style transfer model is used to convert this into a sketch.}
  \label{fig:pipeline}
\end{figure*}

\subsubsection{Extracting Keywords from Reference Images (Figure \ref{fig:pipeline}. (a))}
Based on the findings from our formative study, our pipeline is designed to extract keywords from a provided reference image in four categories: subject matter, action \& pose, theme \& mood, and arrangement. To achieve this, we follow a multi-step process.

To identify the subject matter, action \& pose, and theme \& mood within the image, we employ an image captioning model BLIP-2~\cite{li2023blip2} to generate textual descriptions of the image contents. For a comprehensive understanding of the entire image, we divide it into $3\times3$ segments and generate captions for each segment as well as the whole image. These captioning results are then processed by GPT-4~\cite{openai2023gpt4}, a Large Language Model (LLM), to extract lists of subject matter, action \& pose, and theme \& mood present in the image captions. Prompts used for this are in Appendix \ref{appendix:extractkeywords}.

For the arrangement, we utilize the Segment Anything model~\cite{kirillov2023segment} to generate segments and then identify the top ten prominent segments within the image using the approach from LLM-grounded Diffusion~\cite{lian2023llmgrounded}.
Bounding boxes around these segments provide information about the image's overall structure, such as where the items are placed and where large negative spaces are.

Additionally, for generating the recommendations of the relevant keywords, we use GPT-4, and the prompts used for this are shown in Appendix \ref{appendix:recommendkeywords}.

\subsubsection{Generating Recombinations (Figure \ref{fig:pipeline}. (b))}
When the user selects a set of keywords to generate a new recombination, our system generates a range of options to mix those keywords.

The system first generates three textual descriptions encompassing the selected subject matter, action \& pose, and theme \& mood keywords. Then, it extracts the list of the objects that must be drawn on the image for this description. We use few-shot prompting with GPT-3.5-turbo \cite{openai2020gpt3} for this. The prompt used is in Appendix \ref{appendix:generaterecomb}.
For the arrangement, we developed a layout variator to create layouts similar to the selected image's arrangement while aligned with the generated text description. The layout variator first applies an empirically defined random variation of -50 to 50 pixels on each bounding box component (i.e., x, y, w, h) in the original arrangements. Then, it randomly selects boxes depending on the number of objects that need to be drawn and sorts highly similar layouts first using the similarity metric.
The similarity is calculated by summing the IoU and the complement of the min-max normalized centroid distance between the closest pairs of bounding boxes. Following this similarity, the top five arrangements are utilized for the recombination generation.
The most similar layout is used for generating the image in the initial iteration, and other layouts are used when the user requests more sketches. A few shot prompting with GPT-3.5-turbo is used to map between the arrangements and the objects to create the best image possible. The full prompt is shown in Appendix \ref{appendix:matchlayout}.

However, when the user does not select any arrangement from the references, the system generates a broader range of diverse layout options. A few-shot prompting pipeline using GPT-3.5-turbo generates the three most appropriate layouts for the given text description and object list. This pipeline is built based on the previous work~\cite{lian2023llmgrounded}, and the full prompt for this is in Appendix \ref{appendix:generatelayout}.

Given the textual description and the list of the objects mapped with the generated layout, the system generates images with a layout diffusion model~\cite{li2023gligen}. Following our design goal, the system converts the generated image into a simple line sketch using the U-Net structured style transfer model~\cite{Datasculptor2023styletransfer}.

\subsection{Technical Evaluation}
We evaluated ML-based pipelines, especially for keyword extraction, keyword recommendation, and textual description generation by merging keywords.

\subsubsection{Keyword Extraction Pipeline}
We built a dataset of 100 images with tags categorized by the subject matter, action \& pose, and theme \& mood. We asked 20 people with expertise in design or HCI to annotate five images each.
On average, 5.03, 1.87, and 2.29 keywords in the category of subject matter, action \& pose, and theme \& mood, respectively, were collected per image.

Using this dataset as ground truth, we evaluated the prediction result from the keyword extraction pipeline. Keywords in subject matter and action \& pose categories were matched manually one by one between similar ones. The precision and recall of our pipeline were 94.2\% and 58.2\% in subject matter, and 35.3\% and 51.3\% in action \& pose.
Although some salient keywords in the dataset were missing, the pipeline provided quite accurate keywords in the subject matter.
The predicted action \& pose keywords were not perfectly aligned with the dataset tags, but they were still acceptable on the user side because they were perceived as similar to users even if they were not completely accurate (e.g., for an image of a cat standing straight, our pipeline predicted "stretching arms", while the ground-truth is "dancing").
For theme \& mood keywords, we calculated the cosine similarity of mean embedding vectors~\cite{Wang2020minilm} of ground-truth and predicted result to compare the semantic similarity. This was because for theme and mood, even if words are not exactly the same, there can be many other words that can be accepted as similar. The similarity of the ground truth and prediction was 0.826, which means the keyword extraction model estimates the theme \& mood words quite closely.
Examples of the predictions are presented in the Appendix (Figure ~\ref{fig:examplekeywords}).

\subsubsection{Keyword Recommendation Pipeline} \label{sec:recommendEval}
We evaluated the keyword recommendation pipeline based on whether there was a proper level of similarity between the original keywords and the recommended keywords. This was because it would only be effective if the recommendations were not too similar or irrelevant to the original keywords.

We randomly sampled three to ten keywords from each image-keyword pair in the dataset and made 100 sets of keywords. Then, from the pipeline, we got the recommendations for each set. To verify whether these recommendations have a proper range of diversity, we generated two comparison groups of keywords: the irrelevant group and the synonym group. The irrelevant group consists of random keywords from the dataset, and the synonym group is generated by paraphrasing the keyword in each set. NLTK~\cite{bird2009nlp} and GPT-3.5 were employed to find synonyms.
Then, we used the text embedding~\cite{Wang2020minilm} to calculate the cosine similarity of each group with the original keywords. 

The similarity of the irrelevant and synonym groups to the original keywords was 0.624 and 0.774, respectively, and the recommended keywords had a similarity of 0.696, which is in the middle.
This shows that our recommendations are less similar to original keywords than the synonym group but more similar than the irrelevant group.

\subsubsection{Recombination Generation Pipeline}
The recombination generation pipeline gets a user selection of a set of keywords and generates three different descriptions of the possible image that includes those keywords. As the pipeline aims to provide diverse options, we evaluated the diversity of the description generation model.

Similar to section~\ref{sec:recommendEval}, we built 100 sets of keywords randomly extracted from the dataset. We generated three descriptions using our pipeline for each set, calculated the cosine similarities between those three, and averaged them. Here, we calculated diversity as \(1 - similarity\).
To validate our description generator, we prepared two more description sets, one consisting of explicitly unrelated descriptions randomly acquired from the dataset, and the other one consisting of descriptions that merely paraphrase one of the generated descriptions using paraphraser with T5-based model~\cite{chatgpt_paraphraser}.
The diversity within the random and paraphrased groups was 0.801 and 0.209, respectively, while the generated descriptions from our pipeline show a diversity of 0.395. This indicates that generated output is more diverse than just paraphrasing and less diverse than random ones, which means that the pipeline generates descriptions of a reasonable amount of diversity.

We didn't evaluate the later part of this pipeline, which is about generating images and transforming them into sketches, as we used models from previous research~\cite{lian2023llmgrounded} without any customization or adaptation.

\section{Evaluation}
We conducted a within-subjects comparative study with 16 participants. As our design goals encompassed two steps of the reference recombination -- (1) Finding elements (DG 1 and DG 2) and (2) Recombining elements (DG 3 and DG 4), we first observed how \sysname{} supported each of these steps.
We also evaluated whether \sysname{} eventually improves designers' idea generation results and how it supports the creative process.

\begin{itemize}
    \item RQ1. How does \sysname{} support the two steps of the recombination process---finding elements from the references and recombining elements?
    \item RQ2. Can \sysname{} help users generate better quality and quantity of design ideas?
    \item RQ3. How do users utilize the output of \sysname{} in their ideation process?
\end{itemize}

The baseline system shared a similar interface with \sysname{} but without the key features of \sysname{}---extracting keywords from the reference, suggesting relevant keywords, and generating recombination options. In this baseline system, users could manually leave keyword notes on each reference image, create sketches by specifying layouts and prompts to the image generation model, and use ChatGPT\footnote{OpenAI. (2023). ChatGPT (August 3 Version). https://chat.openai.com/}.
To assess the efficacy of the design of \sysname{}'s features and pipelines rather than the effect of AI functionalities, the same AI functionalities are also included in the baseline system.
After observing prevalent use cases of AI in design processes through recent survey~\cite{designtoolsurvey} and videos~\cite{chatgptdesign, Design2023How, Satori2023Ai}, we included both the language model and the image generation model in the baseline system to simulate real-world scenarios of designers with AI tools. 
The baseline included a model closely aligned with the \sysname{} pipeline to prevent the model performance from affecting the study results. Instead of the GPT models, we provided GPT-3.5-based ChatGPT, and for image generation, we offered the same layout diffusion model as \sysname{}. The screenshot of the baseline interface is presented in Figure \ref{fig:baselinesystem} in the Appendix.

\subsection{Participants}
We recruited 16 participants (10 females, 6 males; age M=24.81 and SD=3.78) through an online recruitment posting.
To determine whether the \sysname{} can handle the challenges found in the formative study with early-stage designers, our participants were set as a group similar to the formative study. We required participants to have a degree in design or art and have participated in at least three different design projects.
11 participants were students majoring in design---5 were at the graduate level, and 6 were at or above the third-year undergraduate level. The other 5 participants have graduated---2 majored in design, 1 minored in design, while others pursued majors in media arts and painting.

All participants also reported having enough sketching skills since we asked them to draw their design ideas during the task. The study was conducted for 2 hours, and we compensated participants with 70,000 KRW (approximately 53 USD).

\subsection{Study Procedure}

\begin{figure*}
  \includegraphics[width=\linewidth]{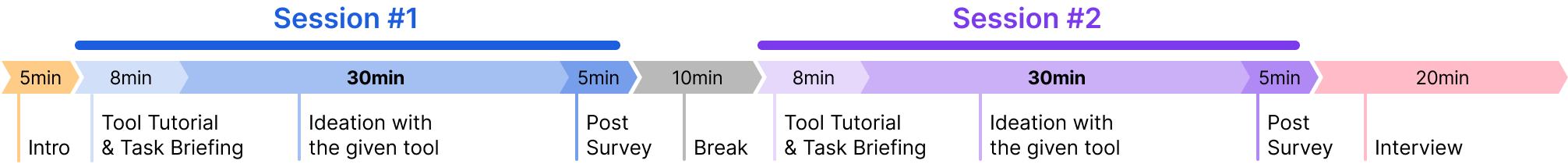}
  \caption{User study process. The 2-hour user study consists of two sessions with different tools, each including a 30-minute ideation phase utilizing the given tool. The order of the tool and the design tasks are counterbalanced. After the two sessions, they had 20-minute semi-structured interviews about their experiences.}
  \Description{User study process. Participants engaged in a 2-hour user study, beginning with a 5-minute introduction outlining the overall process of the study. Then, they underwent two sessions, each including an 8-minute tutorial on the provided tools and task briefing, followed by a 30-minute ideation phase utilizing the designated system. A 5-minute post-survey concluded each session. Participants could have a 10-minute break between the two sessions, and the order of the given tool and the design tasks are counterbalanced. After the two sessions, they had 20-minute semi-structured interviews about their experiences.}
  \label{fig:studyprocess}
\end{figure*}

The whole process of the user study is shown in Figure \ref{fig:studyprocess}.
Participants were asked to perform design ideation tasks twice in two settings: \sysname{} and baseline. The task was to draw an illustration for the cover of a fictional children's book, "Starry Safari: Exploring Alien Jungles" or "A Christmas Dinner in the Underwater World". They were also provided with 10 reference images for each topic. The order of topics and tools was counterbalanced for each participant.

For the first five minutes of each round, participants had a tutorial on the given system and tried it out with sample images to get used to it.
They were then given the topic and the reference images and started ideation using the tool for 30 minutes. If the participants came up with a design idea they wanted to develop further, they sketched it on the paper using a pen. After each round, they completed the post-task survey. Between the two rounds, they could get a 10-minute break. After both rounds, we conducted a 20-minute semi-structured interview to ask about the difference between the two conditions and the effect of the tools on their ideation process. The interview questions are in Appendix \ref{appendix:interviewquestions}.

\subsection{Measures}
The survey after each round included questions about the usefulness of the given system for the different steps of the ideation: organizing the references, discovering useful elements from the reference, exploring multiple ideas, discovering new ideas, and exploring multiple ideas. The survey also included five questions about satisfaction with participants' sketch results regarding overall outcome, quantity, quality, diversity, and creativity. We also had five questions from \cite{aichains} to assess participants' self-perceived experience using the AI system. Participants answered these questions for the image generation feature and ChatGPT after the baseline session, and for the keyword extraction, keyword recommendation, and image/description generating features after the \sysname{} session. Also, the survey included the Creativity Support Index~\cite{csi} and NASA-TLX questionnaire~\cite{HART1988139}.

We also gathered the usage logs (i.e., participant actions with timestamps) to get quantitative metrics for user behaviors. We used this data to calculate the time taken for each sketch, the number of images generated, the number of inputs provided to the image-generating model, etc. Also, every time the participants completed the sketch, the system prompted participants to rate how well the given tool assisted them in producing the idea.

Additionally, we conducted an expert evaluation of the participant's sketches. We recruited two experts with bachelor's degrees in art and had 6 and 1.5 years of experience teaching art each.
We asked them to evaluate two factors in the 7-point Likert scale: (1) the creativity of each sketch and (2) the diversity of ideas within a set of sketches.
We randomly chose three sketches drawn by each participant on each design topic, and a total of 96 sketches (3 sketches x 16 participants x 2 conditions) were evaluated.
The evaluators rated the sketches individually, and for cases of significant score differences (more than 3 points), we asked evaluators to re-evaluate them. While re-evaluating, they were given each other's comments and scores and could choose to change their original score or leave it. They also had to leave comments about their decision as well. There were 9 sketches that required re-evaluation, and all of the conflicts were resolved after one round of re-evaluation. After that, we used the average score for the two evaluators' scores for the result analysis.

\section{Results}
Results showed that \sysname{} helped participants both find and recombine elements for reference recombination. Also, it was shown that users with \sysname{} could generate more design ideas in a given time and perceived their ideas as more creative compared to the baseline. We also found some differences between \sysname{} and baseline regarding how users utilize the tool for their creative process.

\begin{figure}
  \includegraphics[width=\linewidth]{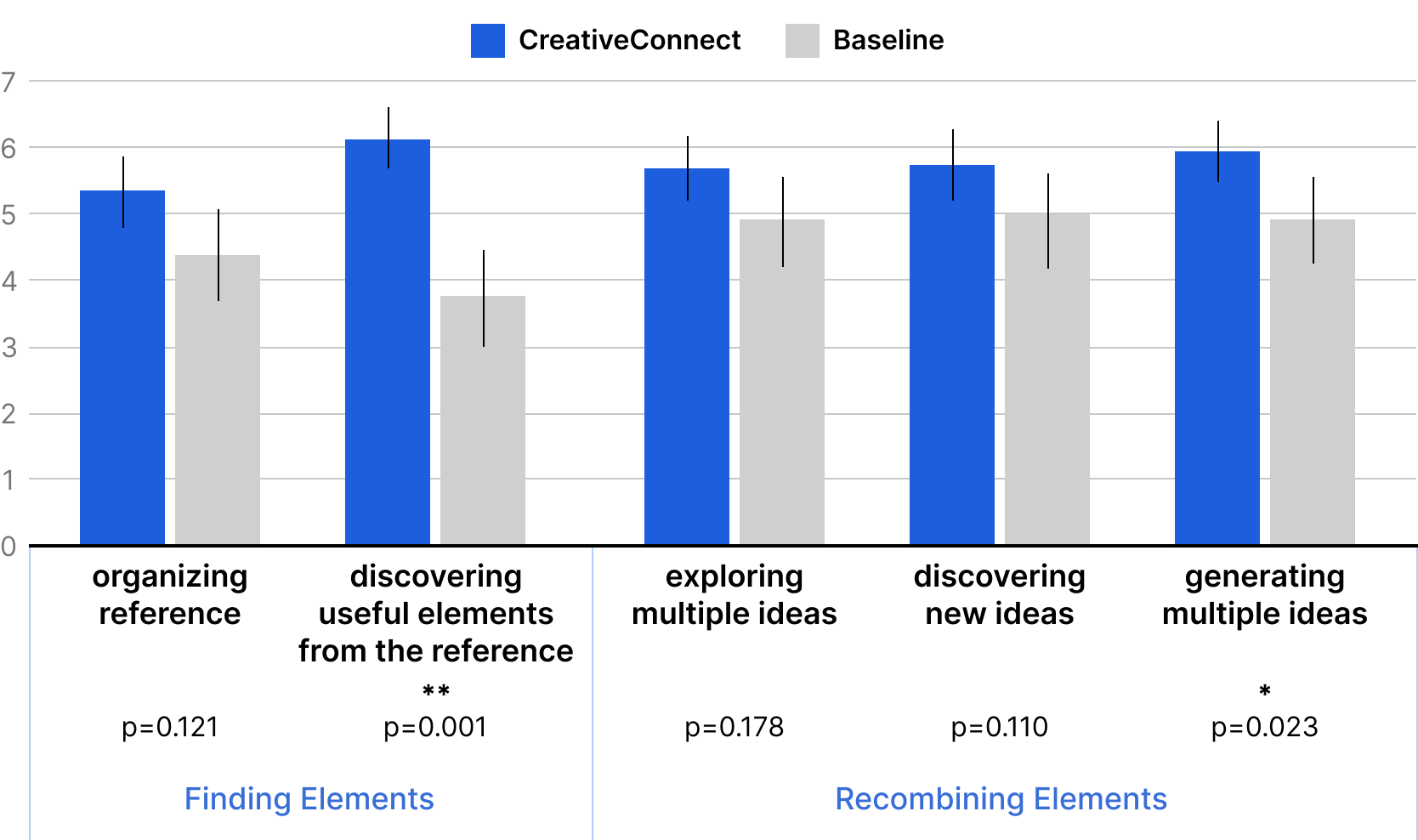}
  \caption{Survey results on the user-perceived efficiency during each recombination step with a 95\% confidence interval. \sysname{} was significantly helpful in discovering elements from the reference image and generating multiple ideas.}
  \Description{Survey results on the user-perceived efficiency during each recombination step with a 95\% confidence interval. The survey has five parts (i.e., organizing references, discovering useful elements from the reference, exploring multiple ideas, discovering new ideas, and generating multiple ideas), which are divided into two recombination steps. \sysname{} is significantly useful in discovering elements from the reference image and generating multiple ideas.}
  \label{fig:survey1}
\end{figure}

\subsection{Support for Different Recombination Steps}
To answer RQ1, we examined survey questions and log analysis results divided into two steps of reference recombination: (1) discovering keywords from the reference images and (2) recombining the found elements into a new concept.
We used a Wilcoxon signed-rank test for all survey questions, as they were ordinal data on a 7-point Likert scale. For the usage log analysis, we conducted a two-sample t-test or two-sample paired t-test to compare between \sysname{} and baseline.

\subsubsection{Finding Keywords from the Reference}
Participants perceived that \sysname{} helped discover valuable keywords from the given reference images. As shown in Figure~\ref{fig:survey1}, participants found out that \sysname{} is significantly more helpful in discovering valuable elements from the reference that can be used for their ideation (M=6.13, SD=1.31) compared to the baseline system (M=3.75, SD=1.98 / p=0.001, W=0.0). 
Regarding how \sysname{} and baseline helped with organizing references, the rating was not significantly different, but with a slightly higher average rating for \sysname{} (Baseline: M=4.38, SD=1.96 / \sysname{}: M=5.31, SD=1.58 / p=0.121, W=23.5).

\begin{figure*}
  \includegraphics[width=\linewidth]{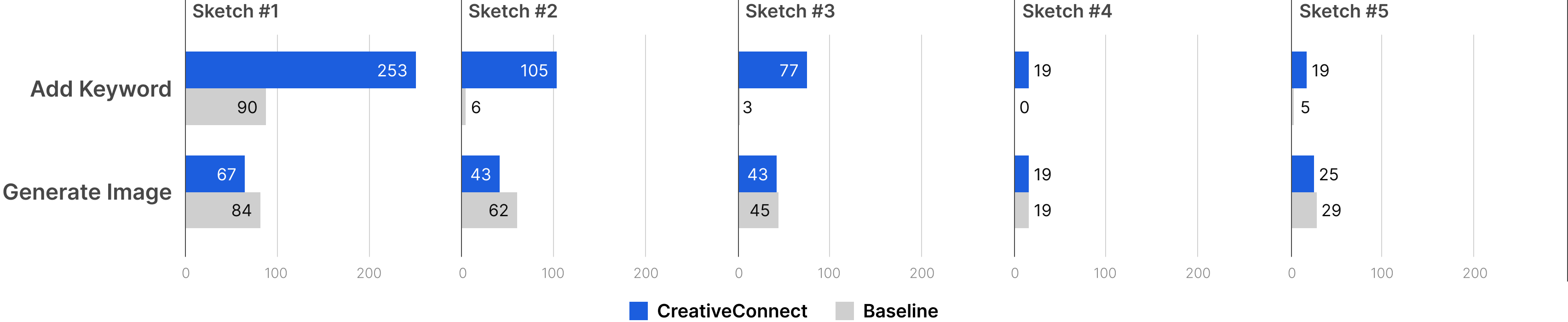}
  \caption{Comparison of the count of two different actions (adding keywords, generating image through generation model) taken to generate each sketch (from the first sketch to the fifth sketch) in \sysname{} and baseline. The results showed that users use the add keyword action more in \sysname{} compared to the baseline, where users only add keywords for the initial sketches. There was no significant difference in the count of generated image action.}
  \Description{The graph shows the comparison of the count of the actions taken to generate each sketch in \sysname{} and baseline. Add keyword action refers to the action where users add a keyword note. Generate image action refers to the point when the user gives input to the image generation model. The results show that users use the add keyword action more in \sysname{} compared to the baseline, where users only add keywords for the initial sketches. There is no significant difference in the count of generated image action.}
  \label{fig:usagelog}
\end{figure*}

Usage logs also showed that \sysname{} effectively encouraged participants to explore and extract different keywords. In comparing the numbers of the keyword notes that participants left in both conditions using a two-sampled paired t-test, participants with \sysname{} added more keyword notes (M=34.69, SD=10.74) compared to the baseline system (M=13.19, SD=10.53 / p<0.0001, t=5.52).
Also, as shown in Figure \ref{fig:usagelog}, participants with baseline typically extracted keywords exclusively during their initial sketch, thereafter relying solely on the previously extracted keywords without actively discovering additional keywords. In contrast, participants using \sysname{} consistently added more keywords throughout the whole process. While they also extracted the most keywords at the beginning, they continued to extract new keywords from references for every new sketch.
One participant (P15) drew all sketches in one go after developing multiple design ideas, instead of sketching immediately after formulating each idea. As we cannot match keyword notes with each specific sketch in this case, this data was omitted from this analysis of actions associated with each sketching instance.
All participants' raw usage log data, including P15, is provided in the Appendix \ref{appendix:rawusagelog}.

\subsubsection{Recombining elements}

The survey's findings indicated that the \sysname{} can be useful for recombining different elements into new design ideas. Participants said that \sysname{} is significantly more helpful (M=5.94, SD=1.34) than the baseline system (M=4.88, SD=1.89 / p=0.023, W=10.5) for them to generate multiple ideas from the collected elements (Figure \ref{fig:survey1}).
However, participants' perception of how much the system helped explore multiple ideas was not significantly different in both conditions, although \sysname{} had a slightly higher average rate (Baseline: M=4.88, SD=1.93 / \sysname{}: M=5.69, SD=1.35 / p=0.178, W=22.0). Also, it was not significant in terms of discovering novel ideas, but the average was slightly higher in \sysname{} (baseline: M=5.00, SD=1.75 / \sysname{}: M=5.75, SD=1.48 / p=0.110, W=19.0).

We also examined how participants used the given image generation model to recombine elements into a design idea. As shown in Table \ref{table:exploration}, there was no significant difference in the number of generated images between the two conditions. However, in exploring diverse recombinations using the model, \sysname{} showed particular advantages, as evident from the unique patterns observed when users interacted with the generation model under two conditions.
Users could provide separate inputs for overall image descriptions and each object using the baseline system. \sysname{} allowed users to select multiple keywords to merge. In both conditions, participants could input multiple phrases together to combine them. We analyzed how diverse phrases are given as a single input into the model.
Out of a total of 347 input sets (202 from the baseline, 145 from \sysname{}), 14 sets (11 from the baseline, 3 from \sysname{}) consisted of only one input, and they were excluded from the analysis since our objective was to compare the semantic similarity between phrases provided to the model together.
For the remaining 333 input sets (191 from baseline, 142 from \sysname{}), we computed the semantic similarity between all pairs of phrases within each input set and calculated the mean and minimum similarity. The mean similarity represents the overall similarities between phrases provided as input together, while the minimum similarity represents the most diverse pairs within the set. Finally, we conducted a two-sample t-test for each metric.

As shown in Table \ref{table:exploration}, the input sets created within \sysname{} showed significantly lower similarity between the keywords (M=0.222, SD=0.094) compared to the sets made within the baseline system (M=0.263, SD=0.166 / p=0.008, t=2.66) when they are calculated based on the minimum similarity. This difference is also similar when they are calculated based on the mean similarity, but it was slightly not significant (Baseline: M=0.356, SD=0.148 / \sysname{}: M=0.330, SD=0.075 / p=0.051, t=1.95).
This means that participants with \sysname{} actively sought to create unique recombinations with greater semantic diversity, ultimately exploring diverse and distinct recombinations compared to the baseline condition.

\begin{table*}[t]
    \begin{tabular}{rrcccccc}
    \toprule
    \multicolumn{1}{l}{}                                                                                   & \multicolumn{1}{c}{}           & \multicolumn{2}{c}{\sysname{}} & \multicolumn{2}{c}{Baseline} & \multicolumn{2}{c}{Statistics} \\ \cmidrule(l){3-8} 
    \multicolumn{1}{l}{}                                                                                   & \multicolumn{1}{c}{}           & mean         & std          & mean          & std          & p               & Sig.         \\ \midrule

    \multirow{2}{*}{\begin{tabular}[c]{@{}r@{}}Image Generation Model Usage\\(Per session)\end{tabular}}         
    & \# of generated image          & 57.06                & 17.91                & 46.69                & 23.52                & 0.119                & -                    \\ \cmidrule(l){2-8} 
    & \# of user inputs to the model & 9.31	              & 4.57                & 10.56               & 4.76                   & 0.468                & -                    \\ \midrule
    \multirow{2}{*}{\begin{tabular}[c]{@{}r@{}}Semantic Similarity\\within Input Sets\end{tabular}} 
    & Semantic Similarity (Mean)     & 0.330        & 0.075        & 0.356          & 0.148         & \textbf{0.051}            & -          \\ \cmidrule(l){2-8} 
    & Semantic Similarity (Min)      & 0.222        & 0.094        & 0.263          & 0.166         & \textbf{0.008}          & \textasteriskcentered{}\textasteriskcentered{} \\  \bottomrule
    \end{tabular}
    \caption{Number of image generation model usage and the semantic similarity between user inputs in \sysname{} and baseline. (-: p > .05, \textasteriskcentered{}: p < .050, \textasteriskcentered{}\textasteriskcentered{}: p < .010, \textasteriskcentered{}\textasteriskcentered{}\textasteriskcentered{}: p < .001)}
    \Description{Number of image generation model usage and the semantic similarity between user inputs in \sysname{} and baseline. The number of generated images and user inputs are not significantly different in the two conditions. Semantic similarity of \sysname{} calculated by minimum value is significantly lower than baseline with a p-value of 0.008. Semantic similarity calculated by mean is also lower in \sysname{} but insignificant with a p-value of 0.051.}
    \label{table:exploration}
\end{table*}

\subsection{Ideation Results}
To answer RQ2, we analyzed the design idea sketches that participants drew during the study session through expert evaluation, usage log, and survey results. Similar to the RQ1, the Wilcoxon signed-rank test was used for survey questions. We used a two-sample t-test for expert evaluation and log analysis results. For pairwise data, such as comparing the number of sketches drawn in each condition by each participant, we conducted a two-sample paired t-test.

\subsubsection{Creativity \& Diversity of the Final Sketches}

\begin{figure*}
  \includegraphics[width=0.75\linewidth]{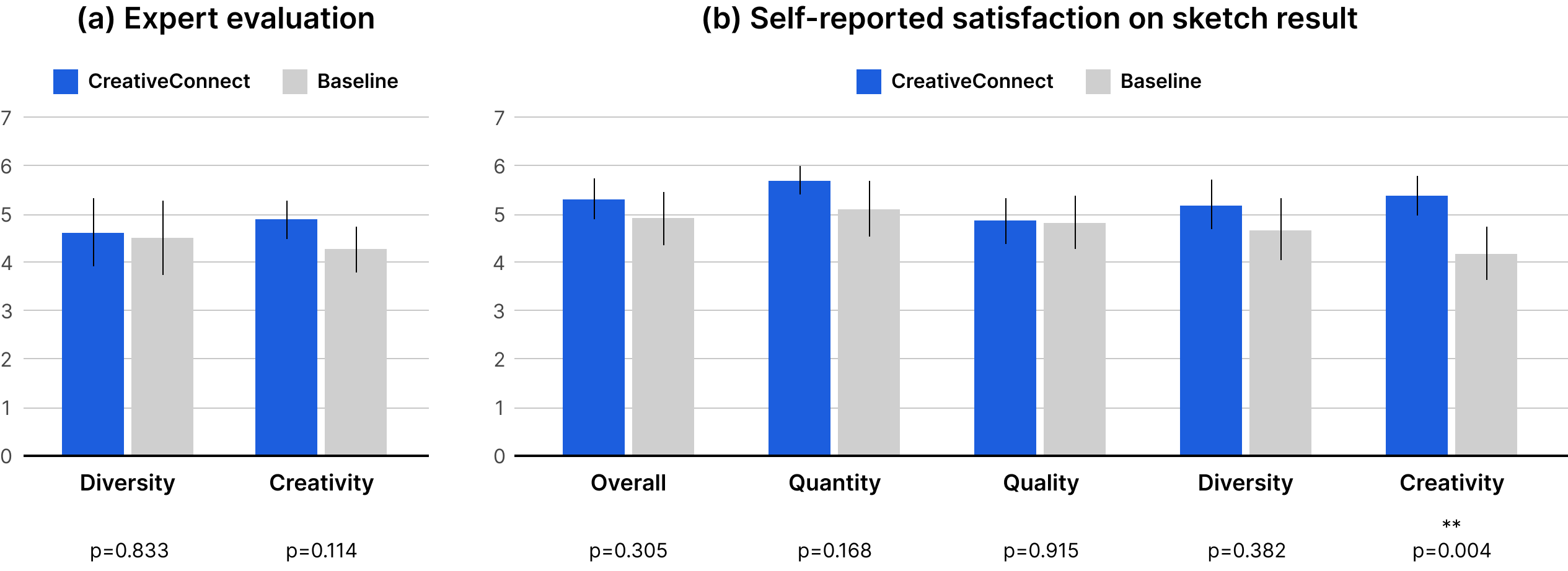}
  \caption{Evaluation results of user-drawn sketches with 95\% confidence interval. (a) Expert evaluation on the diversity and creativity for \sysname{} and baseline condition. (b) Self-reported satisfaction on sketch result in terms of quantity, quality, diversity, creativity, and overall for \sysname{} and baseline condition.}
  \Description{A bar graph shows two evaluation results of the user-drawn sketch with a 95\% confidence interval. (a) Expert evaluation: Statistics (i.e., mean and p-value) of expert ratings of diversity and creativity for \sysname{} and baseline. The diversity of \sysname{} and baseline has no differences in mean, while the creativity of \sysname{} is slightly better but has no significance. (b) Self-reported satisfaction on sketch result: Statistics of self-reported satisfaction ratings regarding quantity, quality, diversity, creativity, and overall in both \sysname{} and baseline. The score of \sysname{} in creativity is significantly higher than the baseline.}
  \label{fig:sketchresult}
\end{figure*}

As shown in Figure \ref{fig:sketchresult} (b), the survey results showed that participants perceived their sketch as more creative when they were using \sysname{} (M=5.38, SD=1.09) compared to the baseline (M=4.19, SD=1.64 / p=0.004, W=0.0). During the interview, 12 out of 16 participants said that they felt they could be more creative with the support of \sysname{} rather than the baseline, especially when they're having a hard time coming up with a new idea in the early ideation stage.
There were no significant statistical differences between the two conditions in terms of other factors, including overall satisfaction, quantity, quality, and diversity of the sketches.

However, as shown in Figure \ref{fig:sketchresult} (a), the expert evaluation does not show a significant difference between the two conditions. The creativity score of the expert evaluation was slightly better in \sysname{} (M=4.854, SD=1.418) compared to the baseline (M=4.344, SD=1.708 / p=0.114, t=1.59), but it was not significant according to the two-sample t-test results. There was also no significant difference in diversity (Baseline: M=4.625, SD=1.607 / \sysname{}: M=4.75, SD=1.418 / p=0.833, t=0.23).
There were possible reasons that expert evaluation was different from the survey results. First, even though the experts were asked to focus on the idea as much as possible, the participants' sketch skills were inevitably reflected in the evaluation, and some of the comments left by the evaluators were actually about the sketch skills. There is also a possibility that deviations according to the design topic may have been affected. In fact, sketches about the topic of underwater Christmas were rated higher on average.

\subsubsection{Efficiency of the Ideation Process}

\begin{table*}[t]
    \begin{tabular}{rcccccc}
    \toprule
        & \multicolumn{2}{c}{\sysname{}}                 & \multicolumn{2}{c}{Baseline}                & \multicolumn{2}{c}{Statistics}              \\ \cline{2-7} 
        & mean                 & std                  & mean                 & std                  & p                    & Sig.                 \\ \hline
    \# of sketch per session                        & \textbf{5.56}                 & 1.63                 & 5.06                 & 1.73                 & 0.041                & \textasteriskcentered{}                    \\ \hline
    time per sketch (min)                           & 5.01                 & 2.87                 & 5.39                 & 3.03                 & 0.403                & -                    \\ \bottomrule
    \end{tabular}
    \caption{Number of sketches drawn by the participants per session and the average time taken for sketches. (-: p > .05, \textasteriskcentered{}: p < .050, \textasteriskcentered{}\textasteriskcentered{}: p < .010, \textasteriskcentered{}\textasteriskcentered{}\textasteriskcentered{}: p < .001)}
    \Description{This table shows the number of sketches the participants drew per session and the average time taken for sketches. The per session average number of sketches from \sysname{} is significantly higher than the baseline.}
    \label{table:quant}
\end{table*}

As shown in Table \ref{table:quant}, the two-sample pairwise t-test result showed that participants came up with more sketches in the same 30-minute ideation session with the support of \sysname{} (M=5.56, SD=1.63) than with the baseline (M=5.06, SD=1.73 / p=0.041, t=2.24). This result indicates that \sysname{} can be helpful for efficient ideation. The interview results also demonstrated that \sysname{} could be useful when they have to come up with a lot of ideas in a limited set of references and time, which is a common scenario in professional design tasks where clients provide references and designers must provide drafts with them.

\subsubsection{Perceived Workload}
As shown in Table \ref{table:quant_merged}, there was no difference between the two conditions regarding the perceived workload. While \sysname{} has additional complications, such as requiring users to specify keywords to give inputs to the image generation model, this does not cause users to feel overwhelmed while performing the task.

\subsection{Impact on User's Creative Process}

\subsubsection{Source of the Inspiration}

\begin{table*}[!t]
    \begin{tabular}{@{}ccrcc@{}}
    \toprule
    \multicolumn{3}{l}{}    & \sysname{}   & Baseline \\ \midrule
    \multirow{5}{*}{\begin{tabular}[c]{@{}c@{}}Source of\\ Inspiration\end{tabular}} &
        \multirow{3}{*}{\begin{tabular}[c]{@{}c@{}}Within\\ the tool\end{tabular}}     
             & Generated image/description & 9       & 5        \\  &                                                                                
             & Recommended keywords        & 1       & -        \\ &                                                                                
             & ChatGPT answers             & -       & 2        \\ \cmidrule(l){2-5} &
        \multirow{2}{*}{\begin{tabular}[c]{@{}c@{}}Outside of\\ the tool\end{tabular}} 
            & Own creativity              & 1       & 1        \\ &
            & Reference images            & 5       & 8        \\ \midrule
    \multicolumn{3}{r}{Avg. tool assistance rating}     & \textbf{5.625}       & 4.563       \\ \bottomrule
    \end{tabular}
    \caption{Number of inspiration sources by category for the most creative sketches chosen by the participants and the average rating of the efficiency of the tool assistance for drawing those sketches. Figure \ref{fig:usageexample} illustrates the example use cases of inspiration within the tool.}
    \Description{This table shows the counts of the inspiration sources for the most creative sketches chosen by the participants and the average rating of the efficiency of the tool assistance for drawing those sketches. Regarding the source of inspiration, the \sysname{} users had more within the tool, especially from the generated images and descriptions. Compared to that, baseline users got more inspiration from outside of the tool, such as reference images. As a result, the average rating on the efficiency of the tool's assistance was higher in \sysname{}.}
    \label{table:usagepattern}
\end{table*}

To investigate how users use the output from \sysname{} and baseline system for generating new design ideas differently, we asked participants to pick one sketch they think is the most creative for each study session and explain how they got the inspiration for it.

\begin{figure*}
  \includegraphics[width=0.95\linewidth]{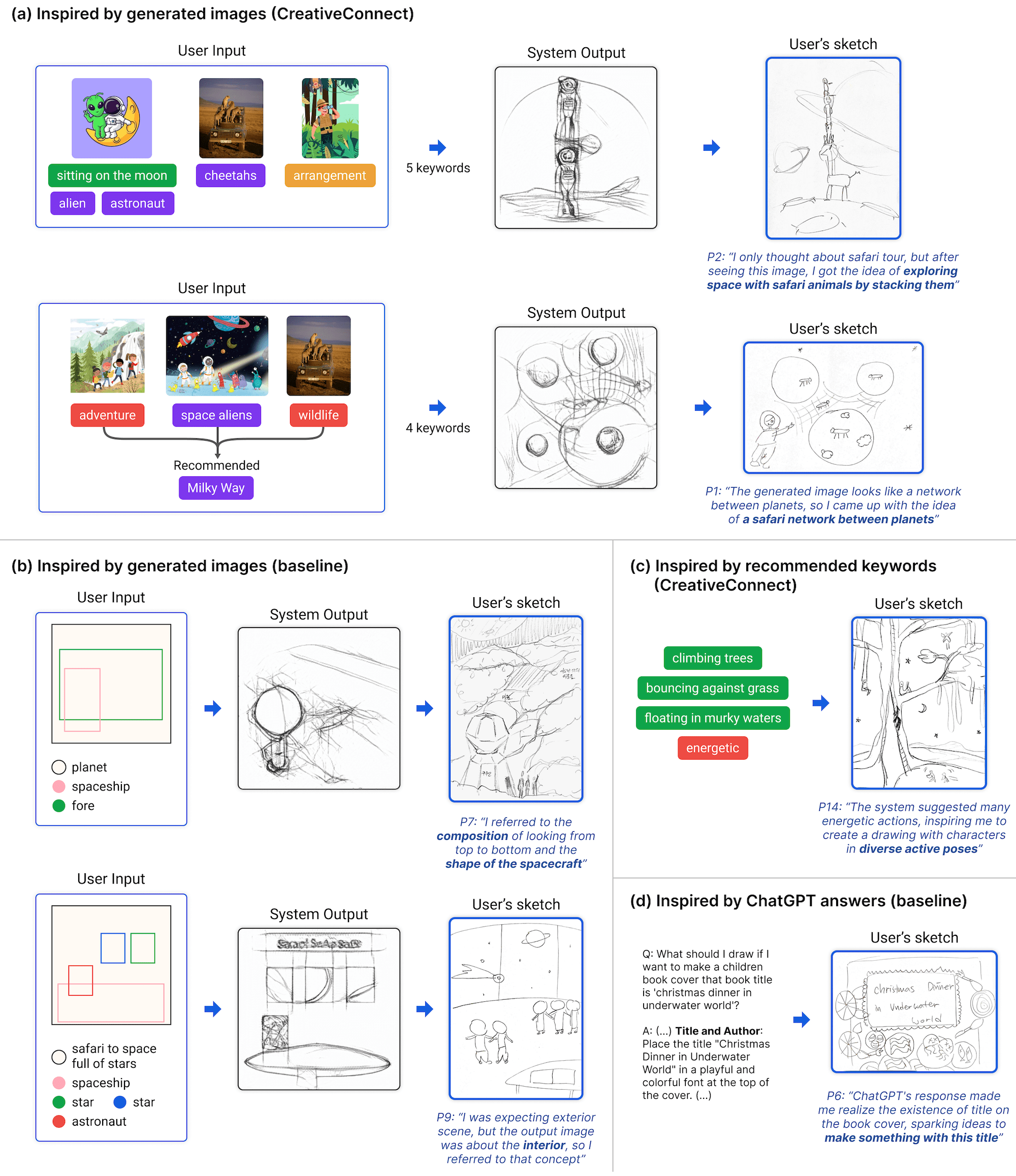}
  \caption{Examples of user input, system-generated output, user-drawn sketches inspired by the system output, and corresponding user quotes. In both conditions, users were inspired by the generated images. However, \sysname{} users were more inspired by the overall concept of the image, while baseline users referred to a specific composition or the detail of the object from the image.}
  \Description{This figure shows six examples of user input, system output usage, and user-drawn sketches. The corresponding user quote is presented in each usage example. (a), (b) Inspired by generated image (\sysname{} and baseline): Two examples of user's keyword input, system-generated image, and inspired sketch. \sysname{} condition's quote is saying, "I only thought about safari tour, but after seeing this image, I got the idea of exploring space with safari animals by stacking them," and "The generated image looks like a network between planets, so I came up with the idea of a safari network between planets." Quote from baseline says, "I referred to the composition of looking from top to bottom and the shape of the spacecraft," and "I was expecting exterior scene, but the output image was about the interior, so I referred to that concept." (c) Inspired by recommended keywords (\sysname{}): a set of recommended keywords from the system and inspired sketch. The quote says, "The system suggested many energetic actions, inspiring me to create a drawing with characters in diverse active poses." (d) Inspired by ChatGPT answers (baseline): ChatGPT answers and sketch inspired by them. The quote says, "ChatGPT's response made me realize the existence of title on the book cover, sparking ideas to make something with this title."}
  \label{fig:usageexample}
\end{figure*}

Five different inspirational sources were found in two conditions. Many participants got their ideas from the generated images or text descriptions in both conditions. In \sysname{} condition, more than half of the participants said that their best ideas are inspired by these generated images or text (Table \ref{table:usagepattern}).
As illustrated in Figure \ref{fig:usageexample} (a), participants utilized keywords from both reference images and recommendations, and merged them using the system. Notably, they got the generated images and tried to reinterpret them in their own way rather than accepting what was drawn there.
Participants with the baseline system were also influenced by the images generated, but the number was slightly less (Table \ref{table:usagepattern}), and how they were influenced was slightly different. They tend to refer to the visual compositions or details of the shapes and apply them to their sketch.
P7 mentioned the reason for this, ``While putting prompts into the image generation model (in the baseline), I already had the concept I wanted. Therefore, I refer to the expression method of it, rather than trying to find something new out of it.''

One noticeable thing is that participants were influenced more by the given reference images when using the baseline. This shows that \sysname{} can make users less directly affected by reference images, ultimately preventing them from fixating on them.
P16 explicitly pointed out this by saying, ``When using \sysname{}, I gave less focus to given images, and as I can expand to a lot of ideas only with a small number of references, I didn't even use all of them.'' P14 mentioned, ``This (baseline) tool feels like a notepad that manages references, so I kept referring to the reference images themselves."

As shown in Figure \ref{fig:usageexample} (d), there was also a participant who got an idea from \sysname{}'s recommended keywords. In baseline, instead of this keyword recommendation feature, they could use ChatGPT, and 2 participants said that they got their inspiration from this.
However, this usage was relatively small (Table ~\ref{table:usagepattern}), mainly because of the challenges of using it for visual tasks. During the interview, participants mentioned difficulties in formulating prompts and leveraging the language-based output for their design.


This difference in sources of inspiration affected the results of the user's rating of how effective the assistance of the tool was. 
We conducted a two-sample pairwise t-test to compare participants' ratings on the tool's usefulness for generating their favorite ideas. The rating was higher in \sysname{} (M=5.63, SD=1.41) compared to the baseline (M=4.56, SD=1.89 / p=0.045, t=2.18) (Table \ref{table:usagepattern}), indicating that users perceived the features of \sysname{} are more helpful in coming up with their best ideas, compared to the baseline system.

The survey results about the perceived experience of using the AI-based system also showed a more specific reason for this helpfulness. As shown in Table \ref{table:quant_merged}, \sysname{} is shown to be significantly better for thinking through what kind of outputs users want to complete for the given task (baseline: M=5.00, SD=1.97 / \sysname{}: M=6.13, SD=1.02 / p=0.045, W=14.5). This shows that participants don't think of the results of the \sysname{}'s image generation model as their final results but more as a guide to thinking about what they want. It leads users to think in diverse ways. P9 mentioned that ``In baseline, the result came out exactly what I thought, so I replicated the output. However, \sysname{} shows me various high-level ways to combine things so I could explore those methods and expand those processes on my own.''

\begin{table*}[t]
    \begin{tabular}{@{}rrcccccc@{}}
    \toprule
    \multicolumn{1}{l}{}                                                                             &               & \multicolumn{2}{c}{\sysname{}} & \multicolumn{2}{c}{Baseline} & \multicolumn{2}{c}{Statistics} \\ \cmidrule(l){3-8} 
    \multicolumn{1}{l}{}                                                                             &               & mean         & std          & mean          & std          & p               & Sig.         \\ \midrule
    \multirow{5}{*}{\begin{tabular}[c]{@{}r@{}}Self-perceived\\experience\\on ML model\end{tabular}} 
     & Match goal    & 5.00         & 1.63         & 4.63          & 1.96         & 0.5805          & -            \\
     & Think through & 6.13         & 1.02         & 5.00          & 1.97         & 0.0454          & \textasteriskcentered{}\\
     & Transparent   & 4.81         & 1.80         & 4.38          & 1.67         & 0.4488          & -            \\
     & Controllable  & 4.75         & 1.95         & 4.06          & 1.84         & 0.2976          & -            \\
     & Collaborative & 5.38         & 1.59         & 4.94          & 2.08         & 0.4809          & -            \\ \midrule
    \multirow{6}{*}{NASA-TLX}                                                                        
     & Mental        & 3.69         & 1.82         & 4.19          & 1.94         & 0.39            & -            \\
     & Physical      & 1.81         & 1.22         & 2.50          & 2.10         & 0.10            & -            \\
     & Temporal      & 2.81         & 1.83         & 3.50          & 2.28         & 0.23            & -            \\
     & Effort        & 3.63         & 1.82         & 3.94          & 2.05         & 0.63            & -            \\
     & Performance   & 5.31         & 1.08         & 5.06          & 1.39         & 0.78            & -            \\
     & Frustration   & 2.63         & 1.93         & 3.50          & 1.75         & 0.14            & -            \\ \midrule
     \multirow{6}{*}{Creativity Support Index}                                                                        
     & Enjoyment            & 5.91      & 1.00     & 5.09          & 1.78         & 0.077 & -                       \\
     & Exploration          & 5.38      & 1.54     & 4.81          & 1.56         & 0.211 & -                       \\
     & Expressiveness       & 5.44      & 1.18     & 4.53          & 1.75         & 0.032 & \textasteriskcentered{} \\
     & Immersion            & 4.69      & 1.99     & 4.69          & 1.82         & 1     & -                       \\
     & Results Worth Effort & 5.47      & 1.27     & 5.25          & 1.71         & 0.591 & -                       \\
     & Collaboration        & 5.19      & 1.25     & 4.41          & 1.71         & 0.016 & \textasteriskcentered{} \\ 
     \bottomrule
    \end{tabular}
    \caption{Survey results of self-perceived experience on ML features, NASA-TLX questionnaire, and Creativity Support Index. (-: p > .05, \textasteriskcentered{}: p < .050, \textasteriskcentered{}\textasteriskcentered{}: p < .010, \textasteriskcentered{}\textasteriskcentered{}\textasteriskcentered{}: p < .001)}
    \Description{Survey results of self-perceived experience on ML features, NASA-TLX questionnaire, and Creativity Support Index. The experience on the ML model is investigated into 5 parts (i.e., match goal, think through, transparent, controllable, and collaborative). \sysname{} is significantly helpful for thinking through. Through the survey on NASA-TLX, \sysname{} and baseline have no significant difference in workload. In the Creativity Support Index, \sysname{} was shown to be significantly preferred in terms of expressiveness with a p-value of 0.032 and collaboration with a p-value of 0.016. There were no significant differences between the two for other criteria.}
    \label{table:quant_merged}
\end{table*}

\subsubsection{Creativity Support Index} \label{section:csi}

According to Table \ref{table:quant_merged}, users prefer \sysname{} significantly more than the baseline regarding expressiveness and collaboration. Still, the other criteria showed no significant difference between the two systems. 
Through the post-interview, we found out that participants felt different types of creativity support in each system. Participants said that the baseline was helpful when they had an overall idea in their mind and wanted to get support for expressing it in the sketch.
On the other hand, participants said that the \sysname{} is helpful for their creativity when they have no idea yet. These differences will be explained in more detail in section \ref{sec:discussion1}.

\section{Discussion}
We propose a novel AI-infused creativity support tool \sysname{}, which assists graphic designers in generating their design ideas by recombining reference images. Based on our findings, we suggest some design implications for future creativity support tools.

\subsection{\sysname{} vs. baseline - Two Different Types of Creativity Support} \label{sec:discussion1}
The results show that \sysname{} successfully supports the early-stage conceptual ideation with reference recombination process by aligning well with the four design goals we derived from the formative study. Participants could easily extract keywords (DG 1) and utilize keyword recommendations as a source of new inspirations (DG 2), leading them to make more keyword notes. Also, they explored diverse keyword recombinations (DG 3), leading them to make more design ideas in a given time. Additionally, they perceived their idea as more creative as \sysname{} provided the output as an incomplete sketch and let participants inject their creativity into it (DG 4).
However, participants didn't feel the difference in the overall degree of creativity support between the two tools. The interviews revealed that this was because \sysname{} and baseline both provided valid creativity support, but in a distinct way based on users' current needs.

In the baseline system, users should specify all the details of the generated image, so they appreciated the transparency and control. The system faithfully reproduced user input by that control, resulting in a final output that closely mirrors the concept in their mind. These generated outputs helped users actualize their existing ideas, more supporting \textit{implementation}~\cite{chung2022artist, chung2021intersection}. Sketch-Sketch Revolution~\cite{fernquist2011sketch} or Framer~\cite{lawton2023tool} had a similar approach to creativity in terms of this.

Conversely, \sysname{} stimulates creativity by providing \textit{inspiration}~\cite{chung2022artist, chung2021intersection}. Instead of requiring users to provide detailed input, \sysname{} accepts keywords and deliberately refrains from exact expression, generating a wider range of outcomes, potentially with serendipity.
How \sysname{} can provide participants with this creative leap can be explained by Cross' descriptive model of creative design~\cite{cross1997descriptive}. The keyword extraction feature actively supports \textit{emergence}, allowing designers to find unrecognized properties of the existing design. The keyword recommendation also supports \textit{mutation}, helping designers to generate new ideas by modifying existing designs partially. P15 metaphorically likened this process to having someone nearby constantly talking with them with fresh variations of ideas. Furthermore, the keyword merging feature enhances \textit{combination}, where new ideas are generated by combining features from existing designs.
Therefore, \sysname{} could be potentially helpful for addressing a common challenge known as ``artist's block'' or ``creative block'', similar to the ``writer's block'' experienced by writers~\cite{hirst1992artists}.
\sysname{} could provide proper support when designers find themselves creatively stuck, breaking creative inertia by sparking novel ideas and opening new creative avenues.

These differences can be valuable design implications for future creativity support tools as designers require different types of creativity support in different stages of the ideation process.
By dynamically adjusting the type of support based on the user's context, such a tool can offer a more personalized and practical creative experience. For instance, when the system detects a user in the exploration phase, it can employ an approach similar to \sysname{}, encouraging the generation of diverse and abstract ideas. Conversely, when the user wants to refine and develop a particular concept, the tool can provide baseline-like features to ensure greater control and fidelity in the generated output. This adaptable approach acknowledges the multifaceted nature of the creative process and supports users with the right tools at the right moment, ultimately enhancing their creativity. Also, integrating those \textit{inspiration} and \textit{implementation} support into a single tool can enable a seamless transition between generating diverse ideas and refining specific concepts, fostering a more iterative and efficient creative workflow.

\subsection{The Role of Low-fidelity Output for Creativity Support}
The post-interview showed that adopting low-fidelity output can facilitate further imagination beyond what the system provided.
We deliberately employ a low-fidelity sketch output in both \sysname{} and the baseline. During the interview, 12 out of 16 participants preferred the sketch output over a complete image, allowing users room for imagination and interpretation. The image converted into a sketch omits small details and retains only the larger forms, generating a large empty space. This emptiness encourages users not just to perceive the generated image but to see it as room for further development and makes users deeply engaged in further ideation.
Some participants even expressed opposition to completed images for the ideation stage, as they believed that an abundance of details in reference images makes them fixated on that specific design idea and hinders them from utilizing the images in their own ideas.
P2 said, ``I usually get completed artworks from Pinterest~\footnote{https://www.pinterest.com/} as a reference, and I found myself unavoidably looking at the unique style of that designer, wanting to replicate it. This time, I liked that I could maintain my own style while exploring different references of concepts.''
Based on our findings, adopting low-fidelity output could be an option when designing creativity support systems to prevent fixation and facilitate the user's creativity in ideation. For example, a design reference tool can dynamically adjust the levels of details of the provided images based on the user's current design stage. When the user wants references for overall concepts, the system can convert reference images to a simple black line drawing or even present it solely as a textual description. Conversely, when the user has determined a specific concept and is exploring different visual details, the system can offer the original images with full details.

\subsection{Generalizability of \sysname{} in Different Context}
\sysname{} is designed to support early-stage designers, such as design students, with a general understanding of the design process but need help with reference recombination. However, our user study revealed some insights applicable to different expertise levels. We observed that participants with limited sketching skills were satisfied more with the baseline system, as it was more aligned with their intentions and suitable for the aid for the actual sketching. Therefore, for users less familiar with artistic expression, an AI tool's output should prioritize alignment with the users' original intent rather than abstraction. Conversely, for experts accustomed to extracting inspiration from references and combining them into their original idea~\cite{bonnardel2005towards}, \sysname{} could serve as a tool for serendipity rather than helping them with the process of keyword extraction and recombination. For example, P16 said that suggested keywords and merged images acted as a prompt to remind them of some aspects initially overlooked. Therefore, features should be redesigned to encourage reflection and creative exploration, such as highlighting the part of the generated images that were not present in existing references but emerged through our system features.

The user study results showed that \sysname{} could also be utilized for other design contexts, such as collaborative projects. According to the CSI survey results (Section \ref{section:csi}), participants indicated that \sysname{} would be significantly helpful for collaborating with other designers. This was because \sysname{} is designed to follow the sequential steps of leaving keyword notes and merging them, and it keeps track of these processes on the mood board and the merging panel. Therefore, participants said that simply showing \sysname{} screen could share their creative processes with other designers, making it easier for them to understand each other's thought processes and quickly reach an agreement on the design direction. One future work direction can be incorporating features of \sysname{} to collaborative mood board tools~\cite{chung2023artinter, koch2020semanticcollage, koch2019may} and studying the benefits of keyword-based recombination features.

\sysname{} could also used for other design domains. 
Our design goals and the feature design of \sysname{} are primarily tailored to the illustration design task, which is predominantly about conveying design topics through visual subject matters and does not usually include other modalities such as text (common in poster or publication design) or motion \& interaction (common in UI/UX and motion graphic design). However, even in other design domains, the recombination process of extracting elements from the reference and recombining them is an effective strategy. To apply the recombination approach to another design domain, we must first identify what elements designers in that domain focus on when looking at references and use those different categories of elements as keywords in the pipeline of \sysname{}.

\section{Limitations and Future Work}
Our work has several limitations that future work can address. In our user study, the ideation tasks were conducted for 30 minutes in each condition, which was shorter than the actual design process. Therefore, it was difficult to observe how the behavior changed over a long time. Future work can be done to incorporate \sysname{} with real-world design projects and see how their behavior patterns differ from lab studies.

Our pipeline generates an image description containing all of the keywords selected by the user as a method of recombination. However, there can be various ways of recombination other than this, such as blending objects or indirectly expressing some keywords through visual details such as colors. Further work can be done on these various recombination methods and how to support them.

As \sysname{} and baseline both leverage generative AI, including LLM and layout diffusion model, the result may be influenced based on users' familiarity with AI. Since this study did not explore those dimensions, future research can examine how creativity supporting tools with AI features may have varying effects depending on the user's knowledge level of AI or prior experiences of using AI.


\section{Conclusion}
This paper proposed \sysname{}, a system designed to support graphic designers in the reference recombination process, allowing them to generate novel design ideas.
Building on our formative study observations, \sysname{} assists users in identifying key elements within reference images. It also provides diverse recommendations for relevant keywords and recombination options. Notably, the low-fidelity sketch-based output of \sysname{} was shown to encourage creativity by enabling further imaginative exploration.
Our user study demonstrated that \sysname{} efficiently supported both steps of finding and recombining elements and helped participants come up with more design ideas and perceive their ideas as more creative than the baseline.
While \sysname{} represents a promising step towards comprehensive recombination support tools for designers, we also suggested an opportunity to expand such systems to address a broader spectrum of design needs and situations.

\begin{acks}
This work was supported by Institute of Information \& Communications Technology Planning \& Evaluation (IITP) grant funded by the Korea government (MSIT) (No.2021-0-01347, Video Interaction Technologies Using Object-Oriented Video Modeling / No.2019-0-00075, Artificial Intelligence Graduate School Program (KAIST))
We thank all of our study participants and the members of KIXLAB for their insightful discussions and constructive feedback.
\end{acks}

\bibliographystyle{ACM-Reference-Format}
\bibliography{main}

\appendix

\section{Technical Details}
\subsection{Prompt: Extracting Keywords from Image Captions}
\label{appendix:extractkeywords}

\begin{framed}
    \footnotesize
    \texttt{
    \\
    \textbf{System Prompt}\\
    You will be provided with multiple sentences to describe an illustration. Your task is to extract a list of Subject matter, Action \& pose, and Theme \& mood.\\
    Subject matters are one-word, describing the specific physical objects, characters, or landscape that the user wants to include in their illustration. Example subject matters include pencil, children, or wave. For subject matters, no adjectives should be included. They should always be a noun.\\
    Actions \& poses are word-level or phrase-level actions that the character or the object in the illustration performs. Example actions \& poses include riding a bus, standing still, or traveling.\\
    Themes \& moods are words not directly present in the illustration, but those that can potentially convey the overall theme or mood of the illustration. Example themes \& moods include imaginative, eco-friendly, or sad. They should be adverbs, preferably one or two words.\\
    If you are provided sentences including some style such as cartoon, illustration, image, or photo, exclude it. For other examples, 'an illustration of a woman sitting at a table' caption is extracted to 'woman', 'table', 'sitting at a table', 'cozy'. The 'illustration' is not contained. Eliminate the changed forms of the same word, such as plurals. Only include roots. For example of 'trees' and 'tree', only include 'tree'."
    \\\\
    \textbf{Examples}\\
    \textbf{User:}
    a card with chinese writing with colorful objects on it\\
    a red and orange background with a blank paper, chinese, pencils, stationery and more\\
    an image of a classroom scene with various supplies\\
    \textbf{Assistant:}
    Subject matter: card, Chinese writing, colorful objects, red and orange background, blank paper, Chinese, pencils, stationery, classroom, supplies.\\
    Action \& pose: \\
    Theme \& mood: education, learning, multiculturalism\\
    \textbf{User:}
    a man and a woman sitting at a table eating food\\
    a woman sitting in a chair in a room with potted plants\\
    a man sitting at a table eating a hot dog\\
    a illustration of a woman sitting at a table in a kitchen with potted plants\\
    an illustration of a woman sitting at a table\\
    an illustration of a refrigerator next to a chair and a table\\
    an illustration of two people sitting at a table with their feet on a chair\\
    a cartoon of a person sitting in a chair in a shower\\
    \textbf{Assistant:}
    Subject matter: man, woman, table, food, chair, potted plants, hot dog, kitchen, refrigerator, feet\\
    Action \& pose: sitting at a table, eating food\\
    Theme \& mood: cozy, heartwarming\\
    \textbf{User:}
    a painting of a white barn in a field of flowers\\
    a painting of a blue house with a tree next to it\\
    a painting of a blue house with a roof and trees\\
    a painting of a window of a house with flowers\\
    a painting of a field of flowers in front of a house\\
    \textbf{Assistant:}
    Subject matter: painting, white barn, field, flowers, blue house, tree, roof, window\\
    Action \& pose: \\
    Theme \& mood: rural, peaceful, nature\\
    \textbf{User:}
    the album cover of the beatles abbey road\\
    a man and a woman standing in front of a car\\
    a man and a woman walking down a street\\
    a group of people walking across a crosswalk\\
    \textbf{Assistant:}
    Subject matter: Beatles, Abbey Road, man, woman, car, street, group, people, crosswalk\\
    Action \& pose: standing in front of a car, walking down a street, walking across a crosswalk\\
    Theme \& mood: urban, nostalgia\\
    }
\end{framed}

\subsection{Prompt: Recommending Relevant Keywords}
\label{appendix:recommendkeywords}

\begin{framed}
    \footnotesize
    \texttt{
    \\
    \textbf{System Prompt}\\
    We are trying to support novice designers' ideation process by semantically combining different parts of illustration references. You will be provided with the topic of the ideation, and multiple keywords users like in the illustrations they found as references. There are three types of keywords: Subject matter, Action \& Pose, and Theme \& Mood.\\
    Subject matters are one-word, describing the specific physical objects, characters, or landscape that the user wants to include in their illustration. Example subject matters include pencil, children, or wave. For subject matters, no adjectives should be included. They should always be a noun. Come up with more than four new keywords for Subject matter.\\
    Actions \& poses are word-level or phrase-level actions that the character or the object in the illustration performs. Example actions \& poses include riding a bus, standing still, or traveling.\\
    Themes \& moods are words not directly present in the illustration, but those that can potentially convey the overall theme or mood of the illustration. Example themes \& moods include imaginative, eco-friendly, or sad. They should be adverbs, preferably one word.\\
    Your task is to expand on the keywords being given, by combining multiple keywords or looking for synonyms that can inspire new creations or ideas. For example, the subject matter "pencil" can be combined with the action \& pose "traveling" to inspire a new action \& pose "writing a diary". You can combine as many keywords at once. Another example is to generate "hair salon" from "hair dryer", "comb", and "scissors". For combinations that result in theme \& mood, make them as abstract as possible. An example is to make "adventurous", "gusty" from "riding on ship" and "tent".\\
    Come up with new keywords for each keyword type with creative combinations. Only use the original keywords provided for creating new keywords. Do not just paraphrase original keywords. Do not suggest similar keywords to the original ones.\\
    Important: Include at least one subject matter for each combination. Subject matter and theme \& mood should be a SINGLE WORD. Combinations among subject matters are highly recommended. New keywords should be \'surprising\' compared to original ones. It means the character of your suggested word should have low similarity.'
    \\\\
    \textbf{Examples}\\
    \textbf{User:}
    Subject matter: camping, tent, tree, animals, Eiffel tower, family\\
    Action \& pose: riding on a bus, riding on a ship\\
    Theme \& mood: playful, imaginative\\
    \textbf{Assistant:}
    Subject matter: bear, sleeping person, safari, cruise\\
    Action \& pose: traveling, setting up camp, dancing jazz\\
    Theme \& mood: adventurous, serene, joyful, romantic\\
    \textbf{User:}
    Subject matter: boy, dinosaur, flower\\
    Action \& pose: watching television\\
    Theme \& mood: fantasy, playful\\
    \textbf{Assistant:}
    Subject matter: wind mill, volcano, movie screen\\
    Action \& pose: exploding strongly, riding a dinosaur, flying away to the sky\\
    Theme \& mood: vast, whimsical, rustic, frenetic\\
    \textbf{User:}
    Subject matter: dreamy scene, boy\\
    Action \& pose: playing with dino toys\\
    \textbf{Assistant:}
    Subject matter: universe, Saturn, astronauts\\
    Action \& pose: imagining adventures, floating on the space, role-playing, daydreaming\\
    Theme \& mood: jolly, imaginative, impactful\\
    \textbf{User:}
    Subject matter: Christmas tree\\
    Action \& pose: dancing around the Christmas tree\\
    Theme \& mood: family-bonding\\
    \textbf{Assistant:}
    Subject matter: Fireplace, wooden sled, Snowman, jazz, piano\\
    Action \& pose: melting, giving present, body-warming\\
    Theme \& mood: jubilant, sparkling, heartwarming\\
    \textbf{User:}
    Subject matter: sea turtles, Christmas tree, marine life\\
    Action \& pose: swimming, dancing around the Christmas tree\\
    Theme \& mood: fantasy, underwater, family-bonding\\
    \textbf{Assistant:}
    Subject matter: Sea horse, Christmas lights, coral, mermaid\\
    Action \& pose: floating on the wave, blinking eye, singing under the sea\\
    Theme \& mood: ethereal, aquatic, charming, panoramic\\
    \textbf{User:}
    Subject matter: kid, cat\\
    Action \& pose: laying on top of a suitcase, playing hide and seek\\
    Theme \& mood: Rustic, vivid, exhilarating\\
    \textbf{Assistant:}
    Subject matter: Birdcage, attic, trunk, blue bird\\
    Action \& pose: jumping on boxes, chasing birds, hiding in a suitcase\\
    Theme \& mood: quaint, mischievous, lively, nostalgic\\
    }
\end{framed}

\subsection{Prompt: Generating Recombinations in Text Descriptions}
\label{appendix:generaterecomb}

\begin{framed}
    \footnotesize
    \texttt{
    \\
    \textbf{System Prompt}\\
    The user wants to draw an illustration, with the assistance of you. You will be provided with multiple keywords users want to include in their illustrations. There are three types of keywords: Subject matter, Action \& pose, and Theme \& mood.\\
    Subject matters are one-word, describing the specific physical objects, characters, or landscape that the user wants to include in their illustration. Example subject matters include pencil, children, or wave. For subject matters, no adjectives should be included. They should always be a noun.\\
    Actions \& poses are word-level or phrase-level actions that the character or the object in the illustration performs. Example actions \& poses include riding a bus, standing still, or traveling.\\
    Themes \& moods are words not directly present in the illustration, but those that can potentially convey the overall theme or mood of the illustration. Example themes \& moods include imaginative, eco-friendly, or sad. 
    They should be adverbs, preferably one word.\\
    Your task is to generate three descriptions of the illustration that the user can draw based on the given keywords. The three descriptions should be significantly different from each other. Each description should include three things: "Caption" and "Objects".\\
    "Caption" is a simple description of the overall image of the description. This should include some objects in the "Objects" list. Keep it concise. Do not make it long and do not include unnecessary adjectives.\\
    "Objects" is a list of the objects depicted in the illustrations, and a short description of them. The objects should be one of the given "Subject matters", or something related to them. You don't have to include all given subject matters. The given action and concept should also be considered for generating the object's detail.
    \\\\
    \textbf{Examples}\\
    \textbf{User:}
    Subject matter: ball, cat, dog\\
    Action \& pose: jumping\\
    Theme \& mood: playful, peaceful\\
    \textbf{Assistant:}
    1.\\
    Scene: In the backyard, cats and dogs play happily with a ball which is their favorite toy.\\
    Objects: [(ball, a brightly colored ball), (dog, a dog jumping with its tongue sticking out), (cat, an agile cat in mid-jump)]\\
    2.\\
    Scene: A beach where a dog is united in its pursuit of a bouncing ball.\\
    Objects: [(ball, a sandy beach ball), (dog, a spotted dog leaping for the ball)]\\
    3.\\
    Scene: In a residential living room, the cat and dog both defy gravity in their jumps to catch the toy.\\
    Objects: [(toy, a small rubber toy with vibrant stripes), (dog, a bouncy terrier flying mid-air), (cat, a Siamese cat also in mid-jump)]\\
    \textbf{User:}
    Subject matter: dinosaur, drink, hat, chair\\
    Action \& pose: relaxing\\
    Theme \& mood: sunny, relaxed\\
    \textbf{Assistant:}
    1.\\
    Scene: A dinosaur wearing a hat, lounging with a drink on a chair under the sun.\\
    Objects: [(dinosaur, a large but friendly looking dinosaur), (drink, a fruity cocktail with a tiny umbrella), (hat, a wide-brimmed straw hat), (chair, a comfortable looking lounge chair)]\\
    2.\\
    Scene: Dinosaur relaxes in a sophisticated environment with a drink and plops a hat on the chair next to it.\\
    Objects: [(dinosaur, a dinosaur in a suit), (drink, a fancy drink in a crystal glass), (hat, a stylish trilby), (chair, a plush velvet chair)]\\
    3.\\
    Scene: A kid with a hat is playing on a picnic mat with a toy dinosaur.\\
    Objects: [(kid, a kid with pink shirts), (dinosaur, a small green toy dinosaur)]\\
    \textbf{User:}
    Subject matter: dog, teeth\\
    Action \& pose: \\
    Theme \& mood: care\\
    \textbf{Assistant:}
    1.\\
    Caption: A caring veterinarian examining a dog's teeth.\\
    Objects: [(dog, a dog sitting calmly), (teeth, a pair of dental tools), (vet, a veterinarian wearing a lab coat)]\\
    2.\\
    Caption: A child diligently brushing their dog's teeth.\\
    Objects: [(dog, a dog lying on its back), (teeth, a toothbrush and dog toothpaste), (child, a child brushing the dog's teeth)]\\
    3.\\
    Caption: A dental hygienist showing a dog owner the correct brushing technique for their pet's teeth.\\
    Objects: [(dog, a dog standing on a dental examination table), (teeth, a dental mirror and toothbrush), (hygienist, a dental hygienist demonstrating the brushing technique)]\\
    }
\end{framed}

\subsection{Prompt: Matching Layout with Objects}
\label{appendix:matchlayout}

\begin{framed}
    \footnotesize
    \texttt{
    \\
    \textbf{System Prompt}\\
    You are an intelligent bounding box matcher. I will provide you with a caption that describes an illustration, a list of the objects that are included in the illustration, and a list of bounding boxes. Your task is to match bounding boxes to each object to make the illustration most balanced and realistic.\\
    Each bounding box is in the format of (object name, [top-left x coordinate, top-left y coordinate, box width, box height]). The bounding boxes are represented as a proportion. The top-left corner has coordinates [0, 0]. The bottom-right corner has coordinates [1, 1]. The bounding boxes should not go beyond the image boundaries.
    \\\\
    \textbf{Examples}\\
    \textbf{User:}
    A realistic image of landscape scene depicting a green car parking on the left of a blue truck, with a red air balloon and a bird in the sky\\
    \relax[air balloon, car, bird, truck]\\
    \relax[0.041, 0.783, 0.442, 0.179], [0.525, 0.699, 0.408, 0.263], [0.261, 0.458, 0.222, 0.264], [0.525, 0.458, 0.279, 0.195]\\
    \textbf{Assistant:}
    \relax[('car', [0.041, 0.783, 0.442, 0.179]), ('truck', [0.525, 0.699, 0.408, 0.263]), ('balloon', [0.261, 0.458, 0.222, 0.264]), ('bird', [0.525, 0.458, 0.279, 0.195])]\\
    \textbf{User:}
    A realistic top-down view of a wooden table with two apples on it\\
    \relax[apple, apple, wooden table]\\
    \relax[0.219, 0, 0.562, 1], [0.402, 0.138, 0.195, 0.195], [0.402, 0.667, 0.195, 0.195]\\
    \textbf{Assistant:}
    \relax[('wooden table', [0.219, 0, 0.562, 1]), ('apple', [0.402, 0.138, 0.195, 0.195]), ('apple', [0.402, 0.667, 0.195, 0.195])]\\
    \textbf{User:}
    A realistic scene of three skiers standing in a line on the snow near a palm tree\\
    \relax[skier, skier, skier, palm tree]\\
    \relax[0.487, 0.131, 0.142, 0.441], [0.661, 0.131, 0.143, 0.441], [0.836, 0.131, 0.142, 0.441], [0.795, 0.613, 0.183, 0.387]\\
    \textbf{Assistant:}
    \relax[('skier', [0.487, 0.131, 0.142, 0.441]), ('skier', [0.661, 0.131, 0.143, 0.441]), ('skier', [0.836, 0.131, 0.142, 0.441]), ('palm tree', [0.795, 0.613, 0.183, 0.387])]\\
    \textbf{User:}
    An oil painting of a pink dolphin jumping on the left of a steam boat on the sea\\
    \relax[dolphin, steam boat]\\
    \relax[0.273, 0, 0.245, 1], [0.032, 0.455, 0.135, 0.420]\\
    \textbf{Assistant:}
    \relax[('steam boat', [0.273, 0, 0.245, 1]), ('dolphin', [0.032, 0.455, 0.135, 0.420])]\\
    \textbf{User:}
    Immersed in his imagination, a boy is indoors enacting a prehistoric tale using four toy dinosaurs.\\
    \relax[dino toys, dino toys, dino toys, boy, dino toys]\\
    \relax[0.250, 0.218, 0.566, 0.563], [0.074, 0.556, 0.137, 0.284], [0.074, 0.76, 0.137, 0.284], [0.659, 0.041, 0.254, 0.134], [0.464, 0.840, 0.195, 0.120]\\
    \textbf{Assistant:}
    \relax[('boy', [0.250, 0.218, 0.566, 0.563]), ('dino toys', [0.074, 0.556, 0.137, 0.284]), ('dino toys', [0.074, 0.76, 0.137, 0.284]), ('dino toys', [0.659, 0.041, 0.254, 0.134]), ('dino toys', [0.464, 0.840, 0.195, 0.120])]\\
    \textbf{User:}
    Two pandas in a forest without flowers\\
    \relax[panda, panda]\\
    \relax[0.114, 0.399, 0.183, 0.441], [0.733, 0.106, 0.150, 0.441]\\
    \textbf{Assistant:}
    \relax[('panda', [0.114, 0.399, 0.183, 0.441]), ('panda', [0.733, 0.106, 0.150, 0.441])]\\
    }
\end{framed}

\subsection{Prompt: Generating Layout based on Image Caption}
\label{appendix:generatelayout}

\begin{framed}
    \footnotesize
    \texttt{
    \\
    \textbf{System Prompt}\\
    You are an intelligent bounding box generator. I will provide you with a caption for an illustration and a list of the objects. Your task is to generate the bounding boxes for the objects based on the caption. The images are of size 512x512. The top-left corner has coordinates [0, 0]. The bottom-right corner has coordinates [512, 512]. The bounding boxes should not go beyond the image boundaries. Each bounding box should be in the format of (object name, [top-left x coordinate, top-left y coordinate, box width, box height]). If an object must appear several times by the provided caption, multiple bounding boxes may be added for the object.
    \\\\
    \textbf{Examples}\\
    \textbf{User:}
    A realistic image of landscape scene depicting a green car parking on the left of a blue truck, with a red air balloon and a bird in the sky\\
    \relax[car, truck, air balloon, bird]\\
    \textbf{Assistant:}
    \relax[('car', [0.041, 0.549, 0.412, 0.31]), ('truck', [0.525, 0.553, 0.408, 0.313]), ('balloon', [0.129, 0.016, 0.283, 0.264]), ('bird', [0.578, 0.082, 0.279, 0.195])]\\
    \textbf{User:}
    A realistic top-down view of a wooden table with two apples on it\\
    \relax[wooden table, apple]\\
    \textbf{Assistant:}
    \relax[('wooden table', [0.039, 0.289, 0.922, 0.422]), ('apple', [0.293, 0.441, 0.195, 0.195]), ('apple', [0.547, 0.441, 0.195, 0.195])]\\
    \textbf{User:}
    A realistic scene of three skiers standing in a line on the snow near a palm tree\\
    \relax[skier, palm tree]\\
    \textbf{Assistant:}
    \relax[('skier', [0.01, 0.297, 0.271, 0.328]), ('skier', [0.543, 0.375, 0.236, 0.308]), ('skier', [0.289, 0.338, 0.242, 0.303]), ('palm tree', [0.789, 0.205, 0.201, 0.49])]\\
    \textbf{User:}
    An oil painting of a pink dolphin jumping on the left of a steam boat on the sea\\
    \relax[steam boat, dolphin]\\
    \textbf{Assistant:}
    \relax[('steam boat', [0.453, 0.439, 0.502, 0.291]), ('dolphin', [0.041, 0.486, 0.369, 0.24])]\\
    \textbf{User:}
    A cute cat and an angry dog without birds\\
    \relax[cat, dog]\\
    \textbf{Assistant:}
    \relax[('cat', [0.1, 0.131, 0.529, 0.632]), ('dog', [0.589, 0.232, 0.412, 0.445])]\\
    \textbf{User:}
    Two pandas in a forest without flowers\\
    \relax[panda]\\
    \textbf{Assistant:}
    \relax[('panda', [0.059, 0.335, 0.414, 0.441]), ('panda', [0.516, 0.338, 0.434, 0.432])]\\
    \textbf{User:}
    Immersed in his imagination, a boy is indoors enacting a prehistoric tale using two toy dinosaurs.\\
    \relax[boy, dino toys]\\
    \textbf{Assistant:}
    \relax[('boy', [0.367, 0.076, 0.301, 0.486]), ('dino toys', [0.15, 0.469, 0.201, 0.254]), ('dino toys', [0.543, 0.475, 0.188, 0.248])]\\
    }
\end{framed}

\subsection{Example Outputs from the Technical Pipeline}
\begin{figure*}
  \includegraphics[width=\linewidth]{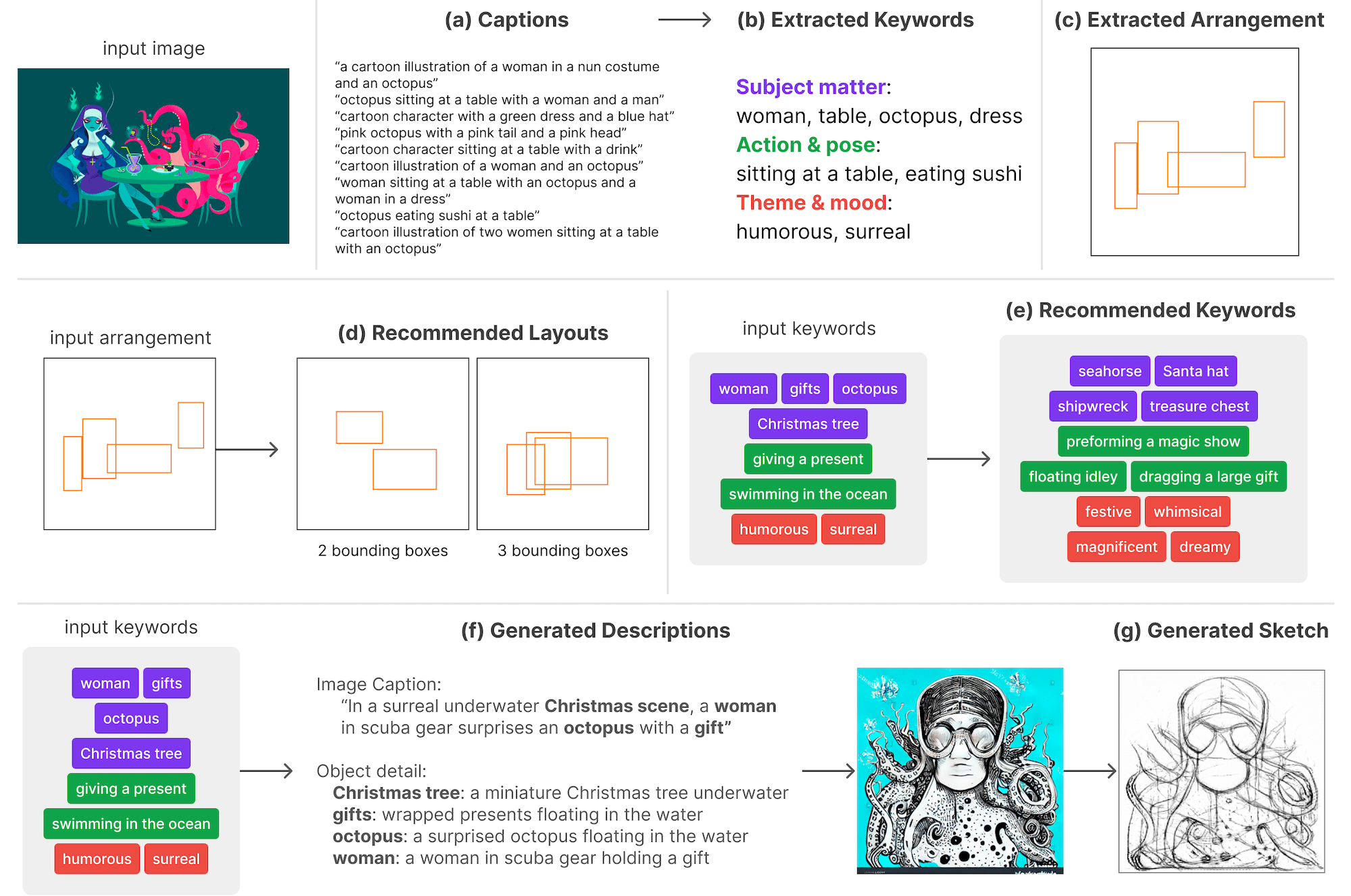}
  \caption{Examples of inputs and outputs given to \sysname{}'s pipelines. From the input image, (a) captions describing a given image and (c) arrangement of an image are acquired. Using captions as input, (b) keywords found in this image are extracted and categorized in subject matter, action \& pose, and theme \& mood. From the (c) extracted arrangement, a layout variator generates two different recommended layouts similar to the original one, each containing 2 and 3 bounding boxes. The keyword recommendation pipeline is used for (e) recommending keywords relevant to the input keywords. Based on the input keywords, (f) descriptions are generated with image captions and the details of objects. The final (g) generated image is created from the descriptions and converted into a sketch style.}
  \Description{Examples of inputs and outputs given to \sysname{}'s pipelines. In this example, an illustration of a witch and octopus having dinner together is given as an input image. From the input image, (a) nine captions describing a given image (e.g., "a cartoon illustration of a woman in a nun costume and an octopus") and (c) arrangement of an image (shown as a bounding box) are acquired. Using captions as input, (b) keywords that are found in this image are extracted and categorized in subject matter (e.g., woman, table), action \& pose (e.g., sitting at a table, eating sushi), and theme \& mood (e.g., humorous, surreal). From the (c) extracted arrangement, a layout variator generates two different recommended layouts similar to the original one, containing 2 and 3 bounding boxes respectively. The keyword recommendation pipeline is used for (e) recommending keywords (e.g., seahorse, performing magic show, festive) that are relevant to the input keywords (e.g., woman, giving a present, humorous). Based on the input keywords, (f) descriptions are generated with image captions ("In a surreal underwater Christmas scene, a woman in scuba gear surprises an octopus with a gift") and the details of objects (e.g., Christmas tree: a miniature Christmas tree underwater). The final (g) generated image is created from the descriptions and converted into a sketch style.}
  \label{fig:technicaloutputexample}
\end{figure*}

\begin{figure*}
    \centering\includegraphics[width=\linewidth]{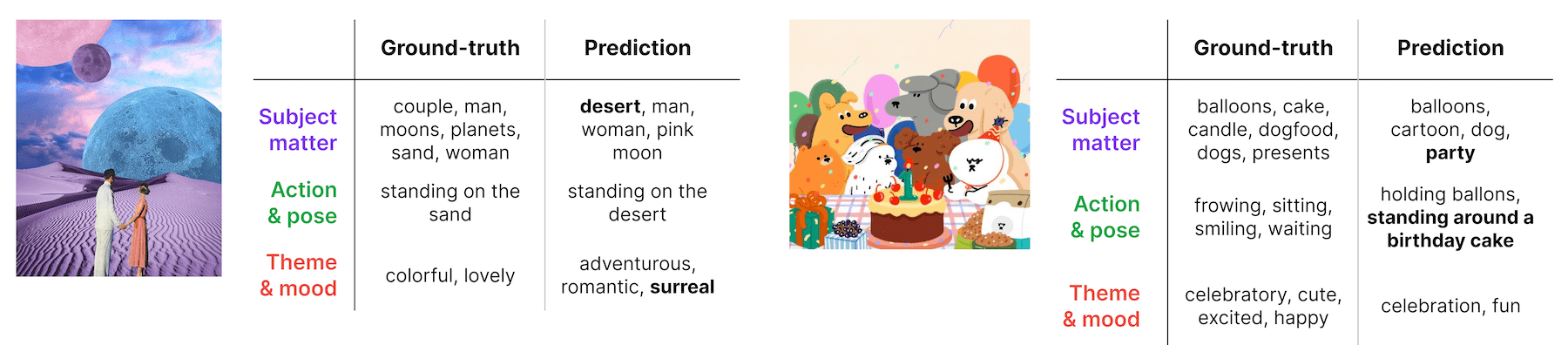}
    \caption{Examples of the reference image with corresponding human-labeled ground truth labels and predicted keywords from the keyword extraction pipeline of \sysname{}. Keywords predicted by our system can often be more descriptive and innovative than the ground-truths, which are highlighted in bold.}
    \Description{Two examples of the reference images with corresponding human-labeled ground truth labels and predicted keywords from the keyword extraction pipeline of \sysname{}. Keywords predicted by our system (right) can often be more descriptive and innovative than the ground truths. For example, there is an image of a couple standing in a purple desert and the big moon in the background. The ground-truth labels only contain the keyword of "sand", but our pipeline predicted more precisely with the keyword "desert". Also, it generates some innovative mood keywords for this image such as "surreal". Another example image is an illustration of a group of dogs celebrating a birthday with a cake. Our pipeline predicted a keyword of "party" or "standing around a birthday cake", which was not shown in the ground-truth data.}
    \label{fig:examplekeywords}
\end{figure*}

Figure \ref{fig:technicaloutputexample} shows some examples of inputs and outputs for three technical pipelines of \sysname{} --- (1) element extraction pipeline, (2) keyword recommendation pipeline, and (3) recombination generation pipeline. Figure \ref{fig:examplekeywords} shows more examples from the keyword extraction pipeline.

\section{User Study}

\subsection{Baseline System Interface}
\begin{figure*}[btp]
  \includegraphics[width=\linewidth]{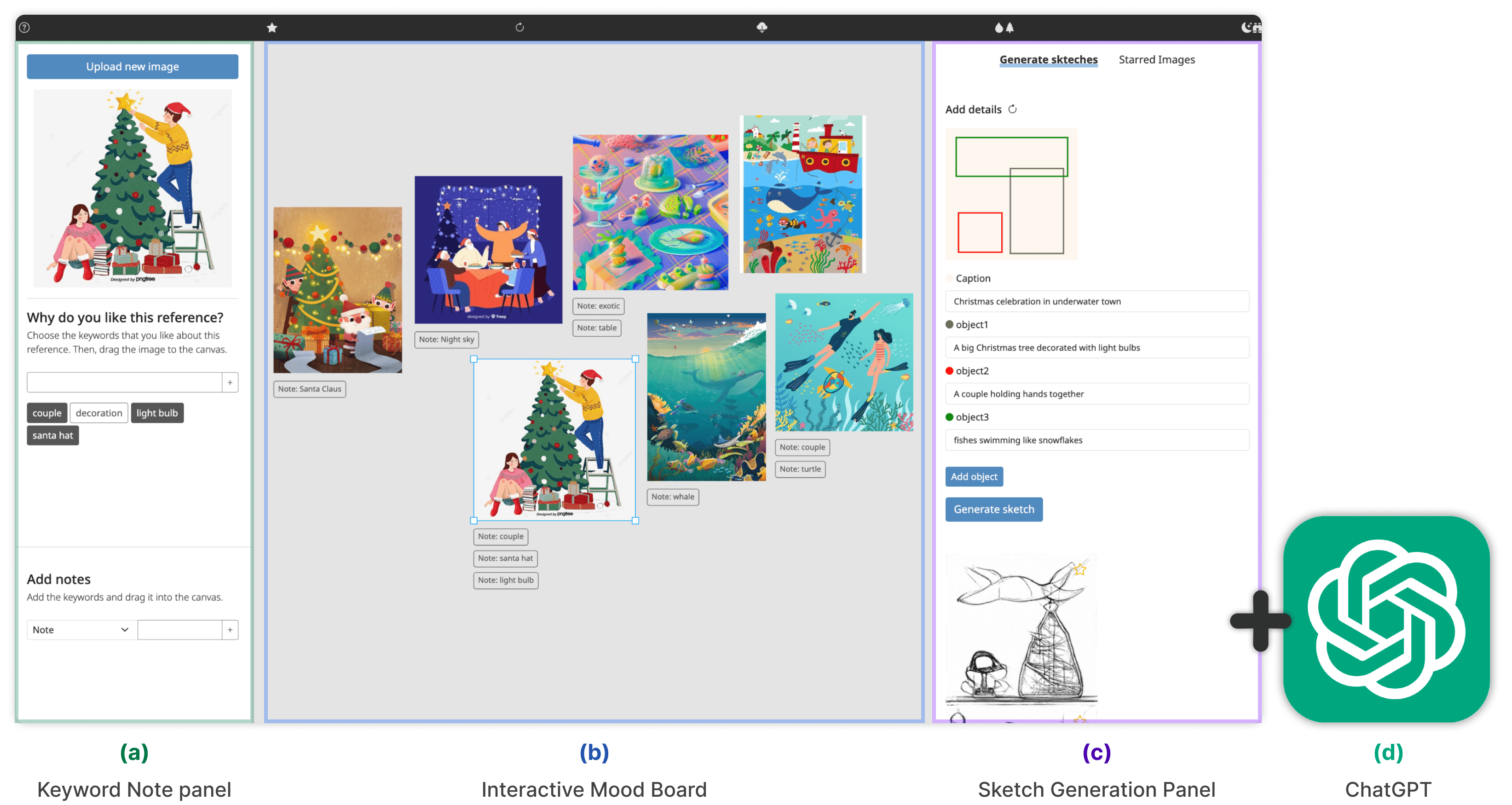}
  \caption{Screenshot of baseline system. (a) \textbf{Keyword Note Panel}: Users can add keywords manually on each image. (b) \textbf{Interactive Mood Board}: Users can organize the reference images on the mood board, along with the added keyword notes. (c) \textbf{Sketch Generation Panel}: Users can configure the overall layout of the generated image by manipulating boxes in the layout controller (beige) on the top of the panel. Additionally, users can provide prompts for the entire image and specific parts. They can specify more image details by clicking the "Add object" button. Users can click the "Generate sketch" button to get the generated sketches. (d) \textbf{ChatGPT}: Users were also provided with ChatGPT on a separate screen.}
  \Description{Screenshot of baseline system. (a) Keyword Note Panel: Users can add keywords manually on each image. (b) Interactive Mood Board: Users can organize the reference images on the mood board, along with the added keyword notes. (c) Sketch Generation Panel: Users can configure the overall layout of the generated image by manipulating boxes in the layout controller (beige) on the top of the panel. Additionally, users can provide prompts for the entire image and specific parts. They can specify more image details by clicking the "Add object" button. Users can click the "Generate sketch" button to get the generated sketches. (d) ChatGPT: Users were also provided with ChatGPT on a separate screen.}
  \label{fig:baselinesystem}
\end{figure*}

Figure \ref{fig:baselinesystem} shows the interface of the baseline system used for the user study.
The baseline system looks similar to the \sysname{}. There is no keyword extraction feature in the left panel, but it allows participants to add keyword notes manually. In the center, there is the same interactive mood board with the \sysname{}, but no keyword suggestion panel exists. The right panel enables users to manually configure the layout and prompts for image generation instead of selecting the keywords to combine. Other features such as mood board interactions (zoom, add/delete images) and saving favorite sketches were provided the same as the \sysname{}. Participants could employ ChatGPT for various purposes other than this interface.

\subsection{Interview questions}
\label{appendix:interviewquestions}

These are the questions used for the semi-structured interview after the two idea generation sessions with baseline and \sysname{} tools.
\begin{enumerate}
    \item Can you share the idea sketch you think is most creative in each topic, and what was the main source of inspiration for those ideas?
    \item Comparing the baseline and \sysname{}, what were the main differences you noticed in the idea generation process?
    \item In each of the three main stages of idea generation—finding reference elements, exploring ideas, and generating sketches—did you find one tool more helpful than the other, and why?
    \item Were there any differences in your typical approach to idea generation when using these tools? If so, how was it different from the usual work process?
    \item Which functionalities were most beneficial in both tools, and in what scenarios were they particularly useful?
    \item Were there any situations or specific sketches where the tools were especially useful or not useful?
    \item In terms of image generation methods, what were the main differences between baseline and \sysname{}, and when did you feel each method was more helpful?
    \item How did you feel about the output in sketch format, and do you think the tool’s effectiveness would differ if outputs were presented as a completed image rather than a sketch?
    \item How did you incorporate the generated images into your final idea sketch?
\end{enumerate}

\subsection{Additional User Study Results: Raw Usage Log}
\begin{figure*}
  \includegraphics[width=\linewidth]{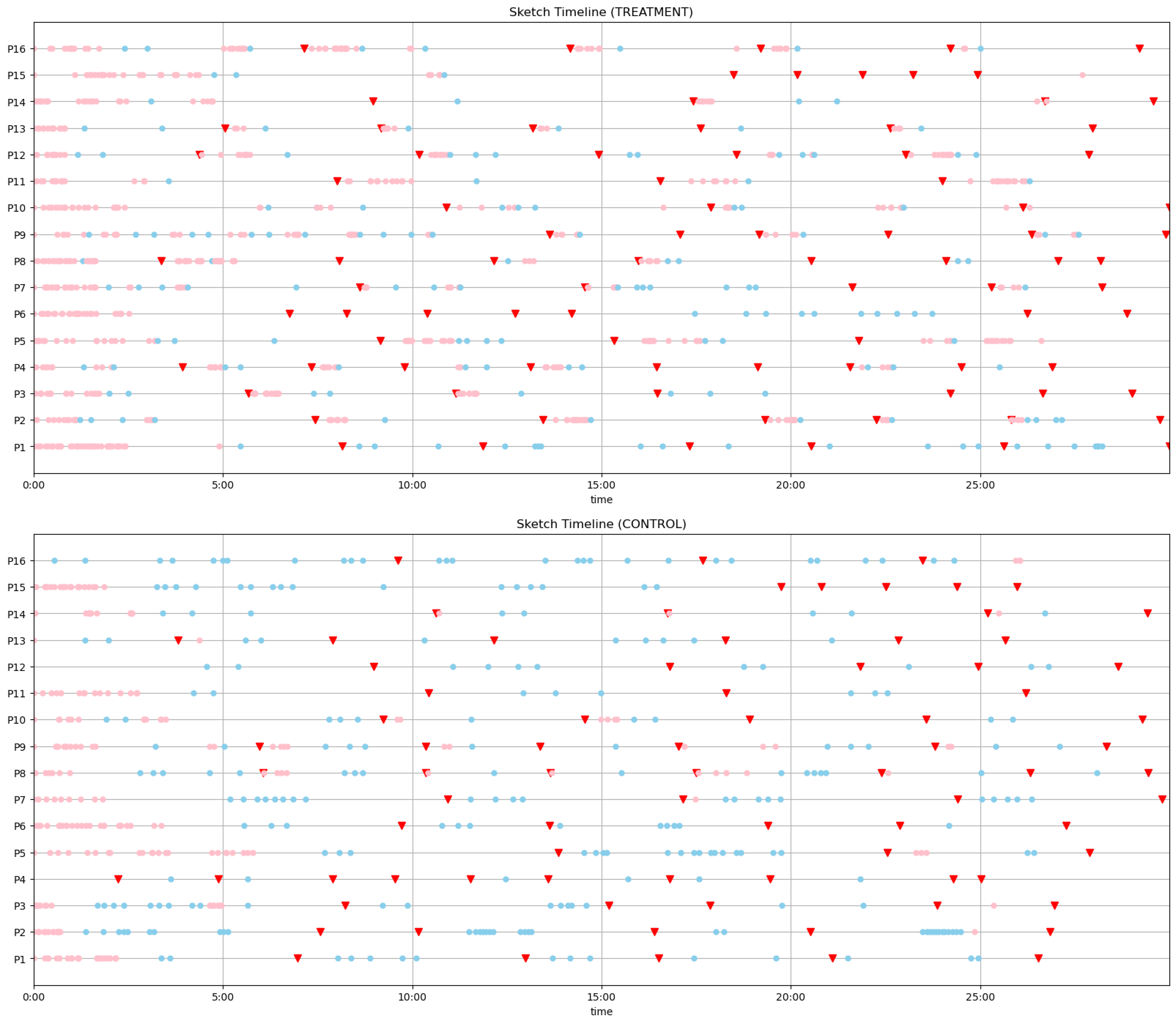}
  \caption{Usage log for all participants in both \sysname{} (Top) and baseline (Bottom) condition. The red triangle indicates the timestamp when the participants complete each sketch. The pink dot is when the participant added new keywords for the reference image. The sky blue dot is when the participant gave input into the image generation model. As shown in the figure, P15 first conducted multiple keyword-adding and image-generation actions, came up with all the design ideas, and then sketched all of them all at once later in the session. During the interview, P15 explained that they intended to focus exclusively on the sketching process, so they decided to jot down half-baked design ideas as memos in a text and draw them collectively. Unfortunately, the collected usage log only records the point when the overall sketch is completed and does not capture the individual instances of writing each memo. Therefore, we could not analyze which actions affected each design idea, so we excluded the usage data of P15 from the analysis of the relationship between action types and each sketching turn (Figure \ref{fig:usagelog}).}
  \Description{Usage log for all participants in both \sysname{} (Top) and baseline (Bottom) condition. The red triangle indicates the timestamp when the participants complete each sketch. The pink dot is when the participant added new keywords for the reference image. The sky blue dot is when the participant gave input into the image generation model. As shown in the figure, P15 first conducted multiple keyword-adding and image-generation actions, came up with all the design ideas, and then sketched all of them all at once later in the session. During the interview, P15 explained that they intended to focus exclusively on the sketching process, so they decided to jot down half-baked design ideas as memos in a text and draw them collectively. Unfortunately, the collected usage log only records the point when the overall sketch is completed and does not capture the individual instances of writing each memo. Therefore, we could not analyze which actions affected each design idea, so the usage data of P15 was excluded from the analysis of the relationship between action types and each sketching turn (Figure \ref{fig:usagelog}).}
  \label{fig:full_usage_log}
\end{figure*}

Figure \ref{fig:full_usage_log} shows the full usage log for all 16 user study participants, showing the timestamps of 3 types of user actions (adding keyword notes, generating images, and completing a design idea sketch).

\label{appendix:rawusagelog}









\end{document}